**Jia-Jun Wu · Bing-Song Zou**

# Hyperon Production from Neutrino–Nucleon Reaction

**Abstract** The neutrino induced hyperon production processes $\bar{\nu}_{e/\mu} + p \to e^+/\mu^+ + \pi + \Lambda/\Sigma$ may provide a unique clean place for studying low energy $\pi\Lambda/\Sigma$ interaction and hyperon resonances below $KN$ threshold. The production rates for some neutrino induced hyperon production processes are estimated with theoretical models. Suggestions are made for the study of hyperon production from neutrino–nucleon reaction at present and future neutrino facilities.

## 1 Introduction

The nature of the lowest hyperon resonances is one of the key puzzling issues in hadron spectroscopy [1]. The lowest iso-scalar excited hyperon $\Lambda^*(1405)$ has been ascribed as a $\bar{K}N$-$\Sigma\pi$ dynamically generated state [2–6], or a pure $|qqq\rangle$ state [7,8], or a $|qqq + qqqq\bar{q}\rangle$ state [1,9]. Where is its iso-vector partner is still not settled. Evidence is accumulating for a low-lying $\Sigma^*$ around 1390 MeV [10–13]. Up to now, most experimental data on these hyperon resonances came from $\bar{K} + N$ reaction. Since these lowest hyperon resonances have masses below $\bar{K}N$ threshold, they can only be observed in multi-hadron final states, suffering complicated strong final state interactions as well as initial state interaction. To avoid such complicated strong initial and final state interactions, the $\bar{\nu}_l + p \to l^+ + \Lambda^*/\Sigma^*$ reactions would provide an ideal clean place to study the low energy $\pi\Lambda/\Sigma$ interaction and hyperon resonances below $KN$ threshold.

On the other hand, in recent years, because of the hot topic study of neutrino oscillation, the neutrino experiments develop rather fast [14]. With increasing intensity of neutrino fluxes, there are many new measurements of neutrino nuclear scattering from various groups, such as MINOS [15], NuTeV [16], SciBooNE [17], ArgoNeuT [18], MiniBooNE [19] at Fermi Lab, and NOMAD [20] at CERN, and K2K [21] in Japan. Although the main purpose of these experiments is to get neutrino-nuclear interaction information needed for the study of neutrino oscillation, these experiments may also provide us a good opportunity to study baryon spectroscopy and baryon structure by the neutrino probes. For the $\bar{\nu}_l + p \to l^+ + \Lambda^*/\Sigma^*$ reactions, the interaction vertices are neutrino-lepton-W boson coupling and s quark-u quark-W boson coupling, which are both well defined in

J.-J. Wu (✉)
Physics Division, Argonne National Laboratory, Argonne, IL 60439, USA
E-mail: wujiajun@ihep.ac.cn

B.-S. Zou
State Key Laboratory of Theoretical Physics, Institute of Theoretical Physics,
Chinese Academy of Sciences, Beijing 100190, China

B.-S. Zou
Theoretical Physics Center for Science Facilities, Institute of High Energy Physics,
Chinese Academy of Sciences, Beijing 100049, China
E-mail: zoubs@ihep.ac.cn

the standard model (SM) [14]. If the quark wave functions of baryons are known, the reaction would be well described. In other words, the $\bar{\nu}_l + p \to l^+ + \Lambda^*/\Sigma^*$ reactions may provide us an excellent place to explore the quark distribution in the hyperon resonances.

With these potential advantages for the study of hyperon spectroscopy and hyperon structure in our mind, here we estimate the $\Sigma^0(1193)$, $\Sigma^{0*}(1385)$, $\Lambda(1115)$ and $\Lambda^*(1405)$ production rates in the $\bar{\nu}_l + p$ reactions with some theoretical models to get some ideas about the possibility to study hyperon resonances from these reactions.

On experimental side, there is no new data from last 30 years. The old data [22–25] only collected less than 100 events for $\Lambda(1115)$ and $\Sigma(1193)$ neutrino production. On theoretical side, several references [26–34] calculated the production of these hyperon resonances: Ref. [26] predicted the $\Lambda(1115)$, $\Sigma(1193)$ and $\Sigma^*(1385)$ neutrino-production; Refs. [27–30] considered the process $\nu + N \to l + Y$(Y=$\Lambda(1115)$, $\Sigma(1193)$); The process $\nu + N \to l + \pi + Y/l + K + N$ was calculated in Ref. [31] where, however, the $\Sigma^*(1385)$'s contribution for $\pi Y/KN$ final state was not included; Ref. [32–34] analyzed the reaction $\nu + N \to l + K + N$ where the vertex of $\Sigma^*(1385)NW$ from $\Delta NW$ by using SU(3) symmetry was assumed for the sub-threshold contribution of the $\Sigma^*(1385)$ isobar. In short, people have predicted the $\Sigma$ and $\Lambda$ neutrino productions [26–30] in the past 40 years, while barely estimated the $\Sigma^*$ and $\Lambda^*$ neutrino productions. Only $\Sigma^*(1385)$ was estimated about 30 years ago [26]. Furthermore, all existing predictions used vector and axial vector current (V–A) approach with vector and axial vector transition form factors. In this method, there are many uncertainties from the various form factor parameters including coupling constants, effective masses, and different types of form factors, such as monopole, dipole and polynomial form factors. Therefore, following the idea of Ref. [9], which used quark wave functions of baryons to study the radiative and strong decay of $\Lambda^*(1405)$, here we use quark model to predict the hyperon resonance neutrino production.

In this paper, three theoretical models are used, namely, the non-relativistic 3-quark model (NR3QM), the V–A with form factors from either some experiment data (EXP) or relativistic 3-quark model (R3QM). Furthermore, although there are no experiment data on $\Sigma^*(1385)$'s weak current, we calculate the $\Sigma^*(1385)$ production by the V–A vertex of $\Sigma^*(1385)NW$ from Ref. [32–34] to compare with our results from the NR3QM.

For the non-relativistic 3-quark model(NR3QM), the baryon wave functions from some non-relativistic 3-quark model and the standard vertices of $\nu_l$-l-W and u-s-W are used to compute the total cross sections of $\bar{\nu}_l + p \to l^+ + \Sigma^0(1193)/\Sigma^{0*}(1385)/\Lambda(1115)/\Lambda^*(1405)$. The SU(3) symmetry breaking effect is studied in this model.

For the V–A approach, the effective form factors are assumed for the p-$\Lambda/\Sigma$-W vertices [28,29], which can be either extracted from the data on weak decays of $\Lambda$ and $\Sigma$ (EXP) [35,36] or from the relativistic 3-quark model (R3QM) [37]. The cross sections of the reactions $\bar{\nu}_l + p \to l^+ + \Lambda(1115)/\Sigma^0(1193)$ are predicted by these two models, which cross check whether the predictions from the non-relativistic 3-quark model are reasonable or not. Actually, our calculation shows that even these two sets of form factor parameters are extracted from the same $\Lambda(1115)$ weak decay experiment data [35], their predictions of the $\Lambda(1115)$ neutrino production rate differ by a factor of 1.25 around $E_\nu = 0.5$ GeV. Obviously, the detection of $\bar{\nu}_l + p \to l^+ + \Lambda(1115)$ reaction will help us to understand the details of the form factors of p-$\Lambda$-W vertex. Furthermore, the precise knowledge of these form factors are extremely important to precisely measure $|v_{us}|$ in the Cabibbo–Kobayashi–Maskawa (CKM) matrix. For the production of hyperon excited resonances, the effective form factors cannot be got directly because there is no experimental data on weak decays of the hyperon excited resonances. Thus EXP method is unable to estimate the neutrino induced production of the hyperon excited resonances.

In short, the neutrino induced hyperon production processes $\bar{\nu}_{e/\mu} + p \to e^+/\mu^+ + \pi + \Lambda/\Sigma$ may provide a unique clean place for studying low energy $\pi\Lambda/\Sigma$ interaction and hyperon resonances below $KN$ threshold, while the $\bar{\nu}_{e/\mu} + p \to e^+/\mu^+ + \Lambda/\Sigma$ may be used to explore the internal structure of $\Lambda/\Sigma$ as well as for the precise measurement of $|v_{us}|$.

In the next section, we present the formulae for calculating the cross sections of neutrino nucleon reactions and branching ratios of $\Lambda(1115)$ and $\Sigma(1193)$ weak decays by three different models. Our results are presented in Sect. 3. In Sect. 4, we give a summary and outlook.

## 2 Formalism

The Feynman diagrams for some typical neutrino nucleon reactions and $\Lambda(1115)/\Sigma(1193)$ weak decays by the W boson exchange are shown in Fig. 1.

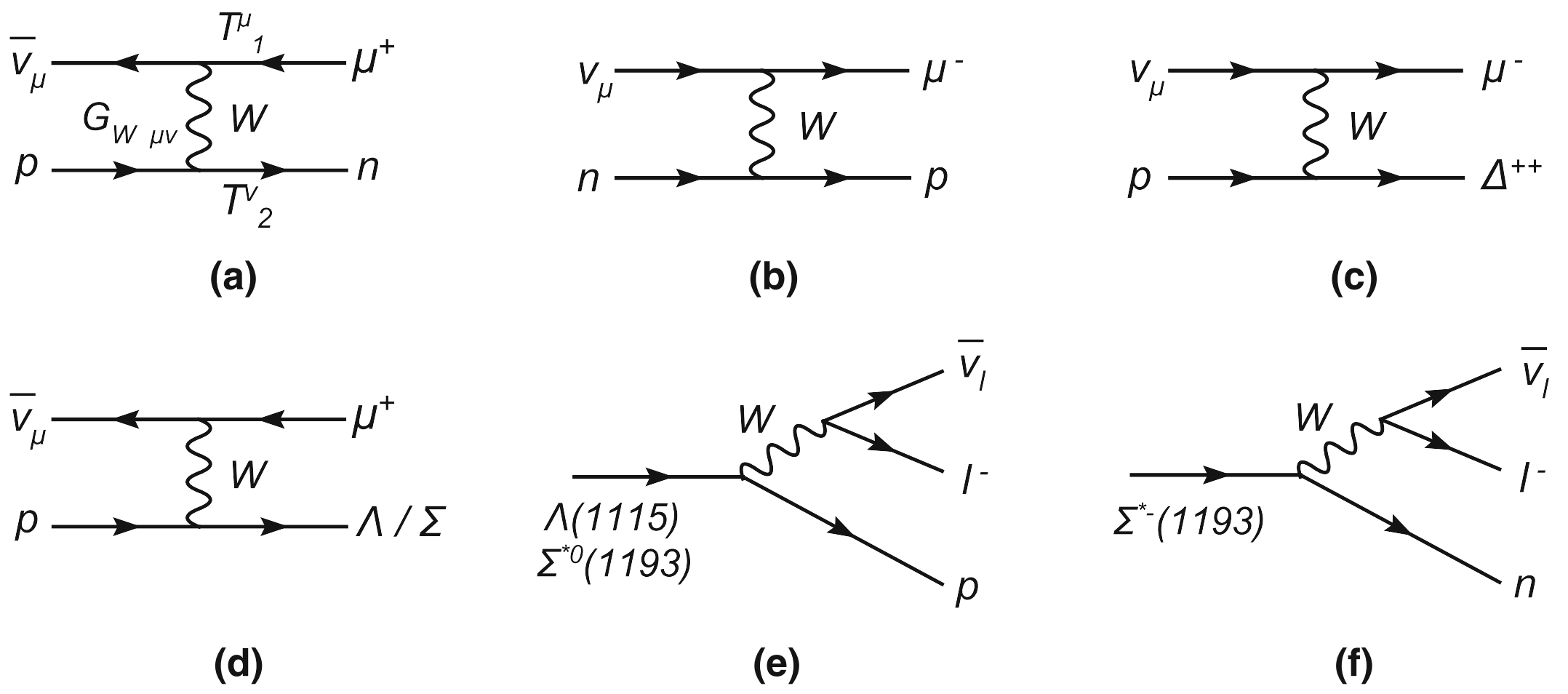

**Fig. 1** Feynman diagrams of typical neutrino nucleon reactions and $\Lambda(1115)/\Sigma(1193)$ weak decays by the W boson exchange. The $l$ in the **e**, **f** stands for $e$ or $\mu$ lepton

There are two vertices: $T_1^\mu$ for the neutrino-lepton-W boson interaction and $T_2^\nu$ for the nucleon-baryon-W boson interaction.

The standard vertex for the neutrino-lepton-W boson in SM is:

$$T_1^\mu = \sqrt{\frac{G_F m_W}{\sqrt{2}}} \left( \bar{l} \gamma^\mu (1 - \gamma^5) \nu_l + h.c. \right), \tag{1}$$

where $G_F$ is the Fermi constant; $m_W$ is the mass of W boson.

For the nucleon-baryon-W boson interaction, $T_2^\nu$, we use NR3QM, and V–A theory with the form factors from EXP [35] and R3QM [37] to perform our calculation. The detailed formulae for $T_2^\nu$ are given in the next two subsections.

Besides the vertices, the propagator of W boson is also needed:

$$G_W^{\mu\nu} = \frac{-g^{\mu\nu} + p_W^\mu p_W^\nu / m_W^2}{p_W^2 - m_W^2}, \tag{2}$$

where $p_W$ is the 4-momentum of the W boson. Because of the heavy mass of W boson, the much suppressed $p_W^\mu p_W^\nu / m_W^2$ term is neglected in the following calculation. With the $T_1^\mu$, $T_2^\nu$ and $G_W^{\mu\nu}$, the amplitudes of the reactions can be easily obtained:

$$\mathcal{M} = T_1^\mu T_2^\nu G_{W\ \mu\nu}. \tag{3}$$

The cross section of $\nu_l(\bar{\nu}_l) + N_1 \to l^-(l^+) + N_2$ can be calculated from the amplitude $\mathcal{M}$:

$$d\sigma = \frac{(2\pi)^4}{2E_\nu} \frac{1}{2} \sum_{s_z^\nu,\ s_z^{N_1}} \sum_{s_z^l,\ s_z^{N_2}} |\mathcal{M}|^2 \delta^{(4)}_{(p_\nu + p_{N_1} - p_l - p_{N_2})} \frac{d^3\mathbf{p}_{N_1} m_{N_1}}{(2\pi)^3 E_{N_1}} \frac{d^3\mathbf{p}_l m_l}{(2\pi)^3 E_l}, \tag{4}$$

where $p_i$ and $\mathbf{p}_i$ are the four and three momentum of particle $i$, respectively; $E_i$ is the energy of particle $i$; $s_z^i$ is the z component of spin of particle $i$. All of these momenta are in the initial baryon rest system.

The formula for calculating the width of $\Lambda \to \bar{\nu}_l + l^- + p$ is:

$$d\Gamma = \frac{(2\pi)^4}{2M_\Lambda} \frac{1}{2} \sum_{s_z^\Lambda} \sum_{s_z^\nu, s_z^l,\ s_z^p} |\mathcal{M}|^2 \delta^{(4)}_{(p_\nu + p_l + p_p)} \frac{d^3\mathbf{p}_p m_p}{(2\pi)^3 E_p} \frac{d^3\mathbf{p}_\nu}{(2\pi)^3 2E_\nu} \frac{d^3\mathbf{p}_l m_l}{(2\pi)^3 E_l}, \tag{5}$$

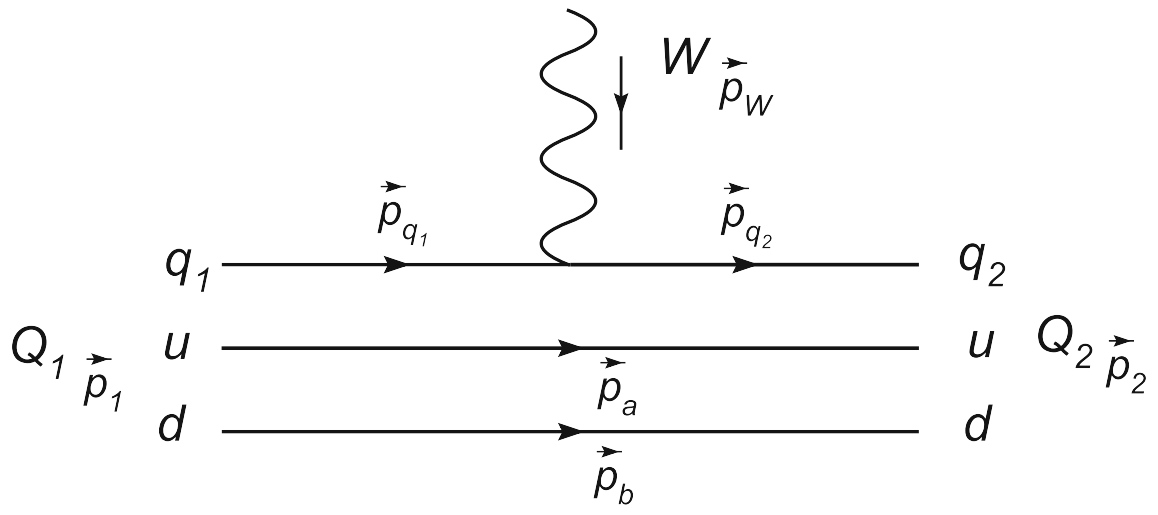

**Fig. 2** Diagram of the nucleon-baryon-W interaction at quark level

### 2.1 $T_2^\nu$ in the NR3QM

In the NR3QM, the wave functions of baryons are taken from Refs. [7,9,39] as follows:

$$|N(939)\rangle = 0.90|N_8^2 S_s\rangle + 0.34|N_8^2 S_s'\rangle - 0.27|N_8^2 S_M\rangle, \tag{6}$$

$$|\Lambda(1115)\rangle = 0.93|\Lambda_8^2 S_s\rangle + 0.30|\Lambda_8^2 S_s'\rangle - 0.20|\Lambda_8^2 S_M\rangle, \tag{7}$$

$$|\Sigma(1193)\rangle = 0.97|\Sigma_8^2 S_s\rangle + 0.18|\Sigma_8^2 S_s'\rangle - 0.16|\Sigma_8^2 S_M\rangle, \tag{8}$$

$$|\Lambda^*(1405)\rangle = 0.90|\Lambda_1^2 P_A\rangle - 0.43|\Lambda_8^2 P_M\rangle + 0.06|\Lambda_8^4 P_M\rangle, \tag{9}$$

$$|\Delta^{++}\rangle = |\Delta^{++} S\rangle, \tag{10}$$

$$|\Sigma^{0*}(1385)\rangle = |\Sigma^{0*}(1385)S\rangle. \tag{11}$$

where the wave functions of $N(939)$, $\Lambda(1115)$, $\Sigma(1193)$, and $\Lambda^*(1405)$ have taken into account the SU(3) breaking effect, while the wave functions of $\Delta$ and $\Sigma^*(1385)$ follow the SU(3) symmetry. The detailed wave functions are given in "Appendix A". In the following calculations, we consider two cases: including the SU(3) breaking effect (named NR3QM-Full) or not (named NR3QM-Single). For the latter, only the first term in Eq. (6–9) is kept, i.e., $|N(939)\rangle = |N_8^2 S_s\rangle, |\Lambda(1116)\rangle = |\Lambda_8^2 S_s\rangle, |\Sigma(1193)\rangle = |\Sigma_8^2 S_s\rangle$ and $|\Lambda^*(1405)\rangle = |\Lambda_1^2 P_A\rangle$. For the $\Delta^{++}$ and $\Sigma^{0*}(1385)$, their wave functions with SU(3) breaking effect are not found in literature, so only predictions based on the NR3QM-Single will be given. The parameters in the NR3QM are the masses of u, d, and s quark, as well as $\omega_3$, the oscillator parameter for the orbital wave function.

The nucleon-baryon-W boson interaction can be described at quark level as shown in Fig. 2, where $Q_1$ and $Q_2$ represent the baryon in the initial state and final state, respectively. The Jacobin momenta used for the initial and final state spatial wave functions are defined as:

$$\mathbf{p}_{1\,\rho} = \frac{\mathbf{p}_a - \mathbf{p}_b}{\sqrt{2}}, \tag{12}$$

$$\mathbf{p}_{1\,\lambda} = \frac{\mathbf{p}_a + \mathbf{p}_b - 2\mathbf{p}_{q_1}}{\sqrt{6}} = \frac{-3\mathbf{p}_{q_1}}{\sqrt{6}}, \tag{13}$$

$$\mathbf{p}_{2\,\rho} = \frac{\mathbf{p}_a - \mathbf{p}_b}{\sqrt{2}}, \tag{14}$$

$$\mathbf{p}_{2\,\lambda} = \frac{\mathbf{p}_a + \mathbf{p}_b - 2\mathbf{p}_{q_2}}{\sqrt{6}} = \frac{-3\mathbf{p}_{q_1} - 2\mathbf{p}_w}{\sqrt{6}}. \tag{15}$$

In the NR3QM, the $T_2^\nu$ can be calculated in the initial nucleon at rest system as follows:

$$\begin{aligned} T_2^\nu(\mathbf{p}_w, s_z^{Q_1}, s_z^{Q_2}) = & \int d\mathbf{p}_a d\mathbf{p}_b d\mathbf{p}_{q_1} d\mathbf{p}_{q_2} \delta_{(\mathbf{p}_a+\mathbf{p}_b+\mathbf{p}_{q_1})} \delta_{(\mathbf{p}_{q_2}-\mathbf{p}_{q_1}-\mathbf{p}_w)} \times \sqrt{\frac{G_F m_W}{\sqrt{2}}} |v_{q_1 q_2}| \\ & \times \sum_{s_z^{q2},\, s_z^{q1}} \langle X^{Q_2},\ s_z^{Q_2},\ \Phi^{Q_2} | \chi^+_{q2,\, s_z^{q2}} \chi_{q1,\, s_z^{q1}} | X^{Q_1},\ s_z^{Q_1},\ \Phi^{Q_1} \rangle \\ & \times \bar{u}_{q_2}(\mathbf{p}_{q_2},\ s_z^{q2}) \gamma^\nu (1-\gamma^5) u_{q_1}(\mathbf{p}_{q_1},\ s_z^{q1}), \end{aligned} \tag{16}$$

**Table 1** Parameters of form factors from fits of hyperon weak decays and relativistic quark model

| Model | Refs. | $f_1(0)$ | $\Lambda_1$ (GeV) | $\Lambda_2$ (GeV) | $g_1(0)$ | $\Lambda_1$(GeV) | $\Lambda_2$(GeV) | $f_2(0)$ |
|---|---|---|---|---|---|---|---|---|
| R3QM | [37] | −1.19 | 0.71 | 0.98 | 0.99 | 0.81 | 1.12 | 0.95 |
| EXP-A | [35,41] | $-\sqrt{3/2}$ | $M_v = 0.97$ | | 0.88 | $M_a = 1.25$ | | 1.19 |
| EXP-B | [35,41] | $-\sqrt{3/2}$ | $M_v = 0.97$ | | 0.90 | $M_a = 1.25$ | | 0.18 |
| $\Sigma^0$(1193)-p-W | | | | | | | | |
| R3QM | [37] | −0.69 | 0.64 | 0.84 | −0.19 | 0.83 | 1.16 | −0.53 |
| EXP | [36] | $-\sqrt{1/2}$ | $M_v = 0.97$ | | −0.231 | $M_a = 1.25$ | | −0.68 |
| $\Sigma^-$(1193)-n-W | | | | | | | | |
| R3QM | [37] | −0.97 | 0.64 | 0.84 | −0.27 | 0.83 | 1.16 | −0.74 |
| EXP | [36] | −1 | $M_v = 0.97$ | | −0.327 | $M_a = 1.25$ | | −0.96 |
| p-n-W | | | | | | | | |
| R3QM | [37] | 1.00 | 0.69 | 0.96 | −1.25 | 0.76 | 1.04 | −1.81 |
| EXP | [38] | 1 | $M_v = 0.84$ | | −1.267 | $M_a = 1.08$ | | −1.855 |

where $s_z^i$ is the z component of spin of particle $i$; $|X^{Q_i}, s_z^{Q_i}, \Phi^{Q_i}\rangle$ is the flavor, spin and spatial wave function of baryon $Q_i$; $v_{q_1 q_2}$ is the CKM element of quark $q_1$ and $q_2$; $\chi^+_{q_i, s_z^{q_i}}$ and $\chi_{q_i, s_z^{q_i}}$ are the creation and annihilation operator of quark $q_i$ with spin $s_z^{q_i}$, respectively; $u_{q_i}(\mathbf{p}_{q_i})$ is the spinor wave function of quark $q_i$. We also give an example of $T_2^\nu$ in "Appendix B".

## 2.2 $T_2^\nu$ in the EXP and R3QM Approaches

At the hadronic level, the baryon-baryon-W boson interaction can be expressed as effective vector and axial vector current as usual. In the V–A theory, the $B$-$N$-W ($B = \Lambda(1115)$, $\Sigma(1193)$ and $N$ = p, n) boson vertex can be written as:

$$T^{\mu}_{2\,BNW} = \sqrt{\frac{G_F m_W}{\sqrt{2}}} |v_{us}| (V^\mu + A^\mu) \tag{17}$$

$$with \quad V^\mu = \bar{B}\left(f_1(q^2)\gamma^\mu - i\frac{f_2(q^2)\sigma^{\mu\nu}q_\nu}{m_B} + f_3(q^2)\frac{q_\mu}{m_B}\right)N + h.c., \tag{18}$$

$$A^\mu = \bar{B}\left(g_1(q^2)\gamma^\mu - i\frac{g_2(q^2)\sigma^{\mu\nu}q_\nu}{m_B} + g_3(q^2)\frac{q_\mu}{m_B}\right)\gamma^5 N + h.c. \tag{19}$$

where $q = p_B - p_N$. According to Refs. [28,29,35,37], the contribution of $f_3$, $g_2$ and $g_3$ are negligibly small. Therefore in the following calculations we only keep the vector, axial-vector and weak magnetism form factors, $f_1$, $g_1$ and $f_2$. They are parameterized slightly different in the literature. As usual, the dipole form for $f_1$ and $g_1$, and constant form for $f_2$ are used:

$$f_1(q^2) = \frac{f_1(0)}{(1 - q^2/M_v^2)^2}, \tag{20}$$

$$g_1(q^2) = \frac{g_1(0)}{(1 - q^2/M_a^2)^2}, \tag{21}$$

$$f_2(q^2) = f_2(0). \tag{22}$$

These form factor parameters are obtained from fits to experimental data of hyperon weak decays [35,36] and are shown in Table 1. In the hyperons' weak decay reactions, the momentum-transfer $q$ is very small. Thus the constant form for $f_2(q^2)$ is good enough to explain the decay data. However, it is not proper for the scattering reactions, since the momentum-transfer would be very large. If the constant form is still applied when the energy of incoming neutrino is larger than 1 GeV, the contribution of weak magnetism term would be much larger than that of vector and axial-vector terms. Furthermore, this constant form would destroy the unitarity seriously. Naturally, the form factor suppressing high momentum-transfer is needed for the $f_2(q^2)$. Unfortunately, the type of form factor and the effective mass are both unable to be got from the data of weak decay. On the other hand, there are some data of $\nu_\mu + n \to \mu^- + p$ and $\bar{\nu}_\mu + p \to \mu^+ + n$, the assumption

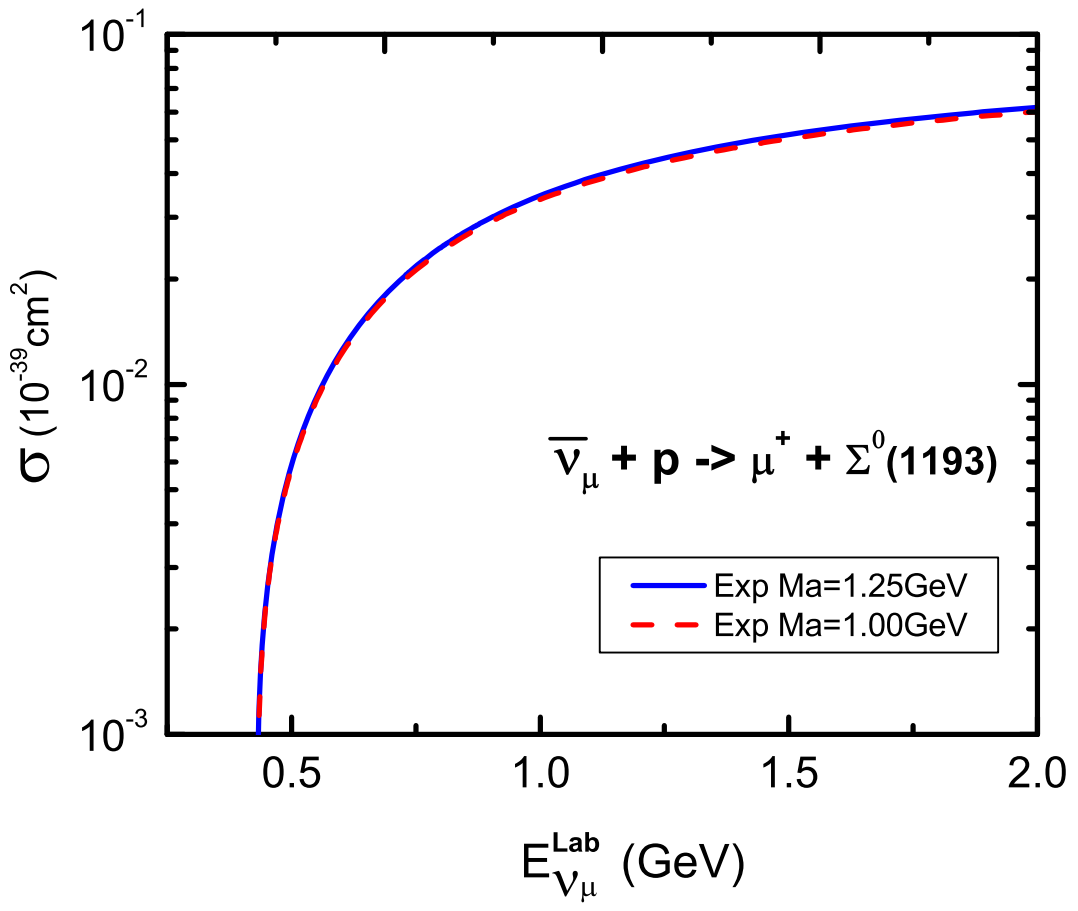

**Fig. 3** (Color online) Comparison of the total cross section of $\bar{\nu}_\mu + p \to \Sigma^0(1193) + \mu^+$ for different axial mass ($M_a$) versus neutrino beam energy in the lab system. The *blue solid, red dashed lines* correspond to $M_a = 1.25$ and 1.00 GeV, respectively. Other parameters are from EXP in Table 1

of same dependence of momentum-transfer of $f_1(q^2)$ and $f_2(q^2)$ can fit these scattering data well as shown in Fig. 4. As a result, we assume that the form factors of $f_2(q^2)/f_2(0)$ and $f_1(q^2)/f_1(0)$ are the same:

$$f_2(q^2) = \frac{f_2(0)}{(1 - q^2/M_v^2)^2}. \tag{23}$$

On the other hand, in Ref. [37], these form factors are obtained from a relativistic quark model calculation with the parametrization form as:

$$f_1(q^2) = \frac{f_1(0)}{1 - q^2/\Lambda_1^2 + q^4/\Lambda_2^4}, \tag{24}$$

$$g_1(q^2) = \frac{g_1(0)}{1 - q^2/\Lambda_1^2 + q^4/\Lambda_2^4}, \tag{25}$$

$$f_2(q^2) = f_2(0). \tag{26}$$

As the same as the EXP model, the $f_2(q^2)$ should be suppressed at high momentum-transfer here:

$$f_2(q^2) = \frac{f_2(0)}{1 - q^2/\Lambda_1^2 + q^4/\Lambda_2^4}, \tag{27}$$

where $\Lambda_1$ and $\Lambda_2$ are the same as them in the $f_1(q^2)$.

The parameters are also listed in Table 1. Kuzmin and Naumov [40] discussed the uncertainties of axial mass. Actually, by our calculation as shown in Fig. 3, the total cross section is not sensitive to the axial mass.

## 3 Results and Discussion

With the formalism and ingredients given in last section, we compute the total cross sections of $\bar{\nu}_\mu + p \to \mu^+ + n$, $\nu_\mu + n \to \mu^- + p$, $\nu_\mu + p \to \mu^- + \Delta^{++}$, $\bar{\nu}_{e/\mu} + p \to e^+/\mu^+ + \Lambda(1115)$, $\bar{\nu}_{e/\mu} + p \to e^+/\mu^+ + \Lambda^*(1405)$, $\bar{\nu}_{e/\mu} + p \to e^+/\mu^+ + \Sigma^0(1193)$, $\bar{\nu}_{e/\mu} + p \to e^+/\mu^+ + \Sigma^{0*}(1385)$ for the neutrino beam energies $E_\nu$ up to 2 GeV, and the decay widths of $\Lambda(1115) \to \bar{\nu}_{e/\mu} + e^-/\mu^- + p$ and $\Sigma(1193) \to \bar{\nu}_{e/\mu} + e^-/\mu^- + N$. For the $\nu N$ reactions, when $E_\nu = 2$ GeV, the momentum-transfer $Q^2 = (p_B - p_N)^2$ is about $0 \sim -5.5\,\text{GeV}^2$, and the three momentum of outgoing baryon is about 0.2–3.2 GeV. It would be the upper limitation for our model since we neglect the boost factor of the quark wave function in the outgoing baryon.

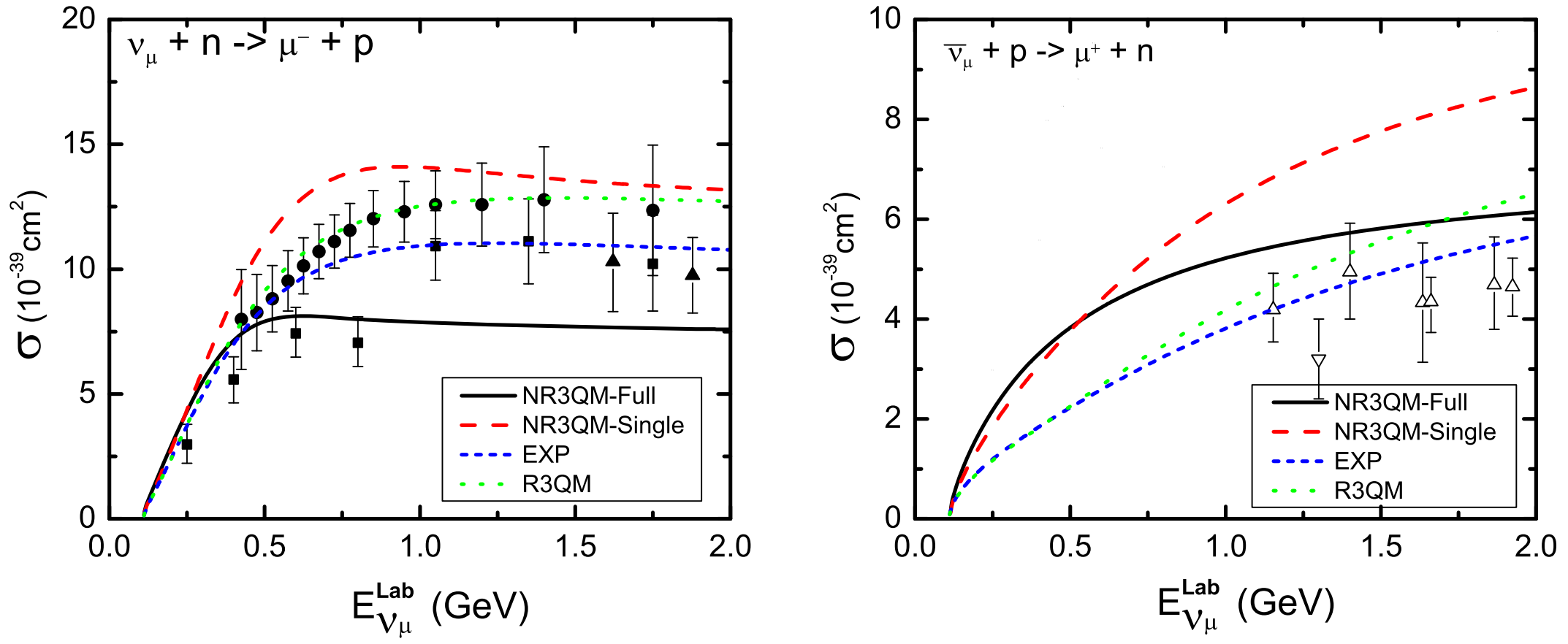

**Fig. 4** (Color online) Total cross sections of $\nu_\mu + n \to \mu^- + p$ (*left*) and $\bar{\nu}_\mu + p \to \mu^+ + n$ (*right*) versus beam energy of $\nu_\mu$ and $\bar{\nu}_\mu$ in the lab system. The *black solid lines* correspond to the nucleon wave function with SU(3) breaking effect. The *red dashed lines* correspond the nucleon wave function of SU(3) symmetry. The *blue short dashed and green dotted lines* correspond to the V–A model with parameters from EXP and R3QM models, respectively. The experimental data are from Refs. [19,42,43,47,48]

## 3.1 Results from the NR3QM

In this model, two types of wave functions for baryons, with and without including SU(3) breaking effect, are used in our calculations. As mentioned in Sect. 2.1, the parameters in the NR3QM are the masses of u, d, s quarks, and $\omega_3$ in the orbital wave functions. By fitting the spectrum of baryons, the values of these parameters are all fixed, $m_u = m_d = 340\,\text{MeV}$, $m_s = 430\,\text{MeV}$, and $\omega_3 = 340\,\text{MeV}$ [9]. Our strategy for this model approach is : (1) to examine whether the results from this model well reproduce the existing experimental data of $\bar{\nu}_\mu + p \to \mu^+ + n$, $\nu_\mu + n \to \mu^- + p$, $\nu_\mu + p \to \mu^- + \Delta^{++}$ and $\Lambda(1115) \to \bar{\nu}_{e/\mu} + e^-/\mu^- + p$, $\Sigma^-(1193) \to \bar{\nu}_{e/\mu} + e^-/\mu^- + n$; (2) using this model with proper parameters to predict the cross section of $\bar{\nu}_{e/\mu} + p \to e^+/\mu^+ + \Lambda(1115)$, $\bar{\nu}_{e/\mu} + p \to e^+/\mu^+ + \Lambda^*(1405)$, $\bar{\nu}_{e/\mu} + p \to e^+/\mu^+ + \Sigma^0(1193)$, $\bar{\nu}_{e/\mu} + p \to e^+/\mu^+ + \Sigma^*(1385)$.

Before showing our results, we want to discuss the status of existing experimental data. The experimental data from the MiniBooNE [19] (solid circle) on the total cross sections of $\nu_\mu + n \to \mu^- + p$ are larger than those from others [42,43] as shown in Fig. 4. A possible reason may be that the MiniBooNE used carbon as the target while others used light target ($H_2$ or $D_2$). The heavier nuclear targets may bring complicated nuclear effects, such as correlations between target nucleons. However, the light target ($H_2$ or $D_2$) would avoid such problem by detecting both muon and nucleon in the final state. Thus, the data from Argonne National laboratory (ANL) [42,44] are more suitable for making comparison with our calculation. We only include the ANL's data for $\Delta^{++}$ production [44]. The cross section of single $\pi^+$ production in MiniBooNE [45,46] by $CH_2$ target is much larger than that of ANL [44] by $H_2$ target with a factor about 10.

In Fig. 4, the total cross sections of $\nu_\mu + n \to \mu^- + p$ and $\bar{\nu}_\mu + p \to \mu^+ + n$ as the function of the neutrino beam energy in the lab system are shown. For the $\nu_\mu + n \to \mu^- + p$ reaction, as shown by the black solid line, if we include the SU(3) breaking effect, the predicted cross sections fit the experiment data (only for square and triangle points) well at E $< 1.0\,\text{GeV}$, but are less than data at higher energies. On the contrary, the calculated cross sections without including the SU(3) breaking effect are a little larger than experimental data with a factor of 1.5 as shown by the red dashed line. For the $\bar{\nu}_\mu + p \to \mu^+ + n$ reaction, the calculation including the SU(3) breaking effect fit the data well while the results without including the SU(3) breaking effect overestimates the cross sections with a factor of 1.3. Furthermore, the blue shot-dashed and green dotted lines are calculated from the EXP and R3QM models, respectively. The predictions of them are both much closer to the data than those of NR3QM. However, there are more parameters which are unable to be applied directly for the predictions of hyperon production, such as axial and vector effective mass. In summary, the predictions of NR3QM models with and without including SU(3) breaking effect are both different from the data within a factor of 2. It is roughly reasonable for estimating the magnitude of cross section since there is no free parameter at all.

In Fig. 5, the total cross section of $\nu_\mu + p \to \mu^- + \Delta^{++}$ vs the beam energy of anti-neutrino is shown. The experimental data are from the ANL [44]. Other groups used multi-nucleons target and their results may not be suitable for comparing with our results directly. Because we do not have the $\Delta^{++}$ wave function with SU(3) breaking effect, the results are calculated by the SU(3) symmetry assumption. The results are still a little larger than experimental data. Note that the pure contribution of $\Delta$ in the $\nu_\mu + p \to \mu^- + \pi^+ + p$ reaction from Ref. [51] (the dashed line in Fig. 5 of Ref. [51]) is consistent with ours, and it is also larger than the ANL data. They used V–A model and the axial coupling $C_5^A(0) \sim 1.15$ from the Goldberger–Treiman relation. The gap between predictions and experimental data indicates that naive three quark model without SU(3) breaking effect may not be good enough to describe $\Delta^{++}$'s structure precisely. Furthermore, the non-resonant contribution is not included in our calculation, while it is important here as discussed in the Refs. [49–53]. The non-resonant contribution might make the cross section even larger, as shown in Ref. [51]. It is definitely a source of the systematic uncertainties in our model. By these considerations, our model is too simple to describe the experimental data precisely. However, we want to point out that our aim is not to find a precise model to describe existing data with a lot of parameters which are not universal for all channels. What we want is to give a reasonable estimation of hyperon neutrino production rates in magnitude. Thus our NR3QM model satisfies this request, since there is no free parameter and the difference between predictions and existing data is less than a factor of 2.

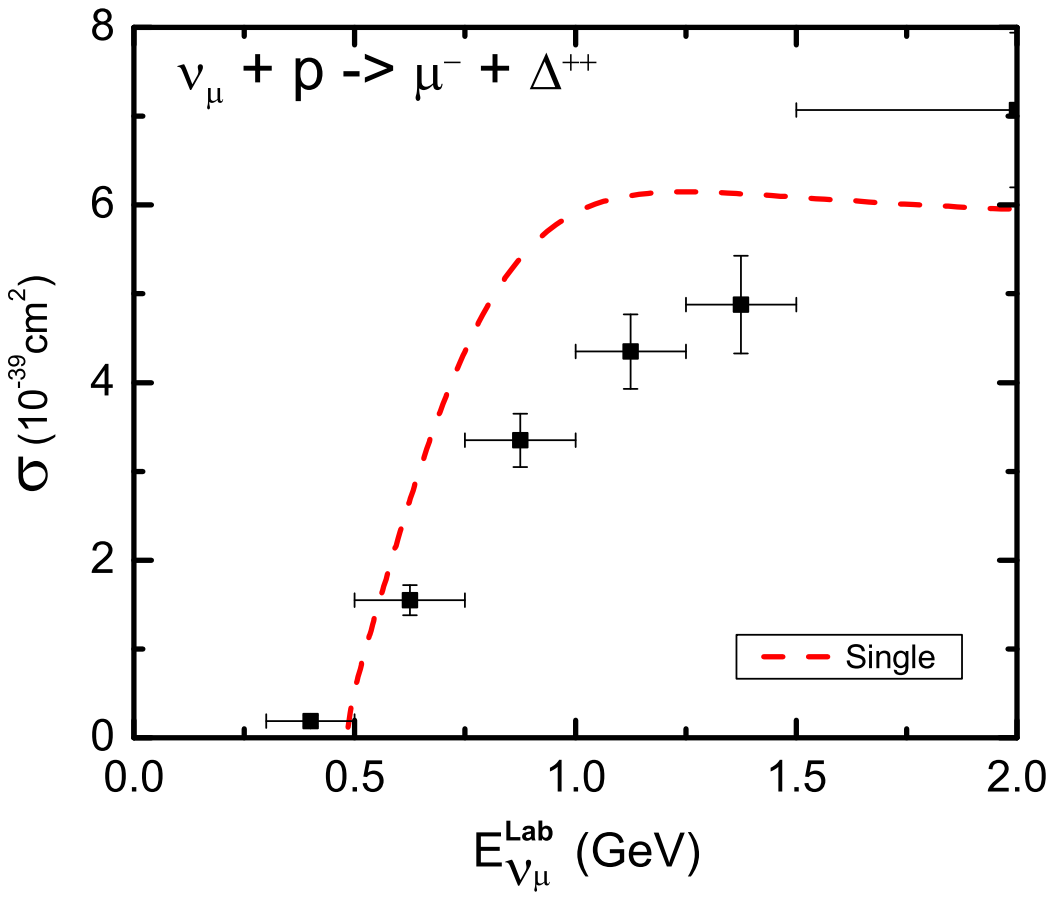

**Fig. 5** Total cross section of $\nu_\mu + p \to \Delta^{++} + \mu^-$ as function of energy of $\nu_\mu$ in the lab system by using the proton and $\Delta$ wave function from SU(3) symmetry. The experimental data are from Ref. [44] for $\nu_\mu + p \to \pi^+ + p + \mu^-$

In Table 2, we list the branching ratio of $\Lambda(1115) \to \bar{\nu}_{e/\mu} + e^-/\mu^- + p$ and $\Sigma^-(1193) \to \bar{\nu}_{e/\mu} + e^-/\mu^- + n$ in the NR3QM (NR3QM-Full and NR3QM-Single) approaches, compared with results from other approaches given in the next subsection and data. And the results in the two cases are slightly larger than the experimental data. And for the $\Sigma^0(1193) \to \bar{\nu}_{e/\mu} + e^-/\mu^- + p$, because of the large decay width of $\Sigma^0(1193) \to \Lambda(1115) + \gamma$, the branching ratio of its weak decay is very small and not be able to be measured yet.

With the comparison between the results of NR3QM and the existing data on the total cross sections of $\bar{\nu}_\mu + p \to \mu^+ + n$, $\nu_\mu + n \to \mu^- + p$, $\nu_\mu + p \to \mu^- + \Delta^{++}$ and the branching ratios of $\Lambda(1115) \to \bar{\nu}_{e/\mu} + e^-/\mu^- + p$ and $\Sigma^-(1193) \to \bar{\nu}_{e/\mu} + e^-/\mu^- + n$, we find that with and without SU(3) breaking effect both fit all data reasonably well. Then we can make predictions for the corresponding $\Lambda(1115)$, $\Lambda^*(1405)$, $\Sigma(1193)$ and $\Sigma^*(1385)$ production rates as shown in Figs. 6 and 7. At $E_\nu = 2.0\,\text{GeV}$ in the lab system, the predicted total cross section for $\bar{\nu}_{e/\mu} + p \to e^+/\mu^+ + \Lambda(1115)$ is $(0.14 \sim 0.28) \times 10^{-39}$ cm$^2$, for $\bar{\nu}_{e/\mu} + p \to e^+/\mu^+ + \Lambda^*(1405)$ is $(8 \sim 11) \times 10^{-42}$ cm$^2$, for $\bar{\nu}_{e/\mu} + p \to e^+/\mu^+ + \Sigma^0(1193)$ is $(0.018 \sim 0.039) \times 10^{-39}$ cm$^2$, and for $\bar{\nu}_{e/\mu} + p \to e^+/\mu^+ + \Sigma^{0*}(1385)$ is $(\sim 10) \times 10^{-42}$ cm$^2$. In Ref. [22], there are 13 events for $\Lambda(1115)$ and 2 events for $\Sigma^0(1193)$, and the cross section of $\Lambda(1115)$ production is $0.13^{+0.09}_{-0.07} \times 10^{-39}$ cm$^2$ at the energy of anti-neutrino from 1 to 3 GeV; and in Ref. [23] with $E_{\bar{\nu}} = 0.5$–10 GeV, they got 83 events for $\Lambda(1115)$ and seven events for $\Sigma^0(1193)$ after correction, the corresponding cross sections are $(0.179 \pm 0.045) \times 10^{-39}$ cm$^2$ and $(0.016 \pm 0.01) \times 10^{-39}$ cm$^2$, respectively.

These existing experimental data for hyperon are well consistent with our predictions. The predicted total cross section for $\Lambda(1115)$ is much larger than that for $\Sigma^0(1193)$, while their $J^P$ are both $\frac{1}{2}^+$. The reason is that the flavour wave functions of them are different since the $\Lambda(1115)$ is singlet while $\Sigma^0(1193)$ is one of the triplet in the baryon octet. When u quark of the initial proton transfers to the s quark of the final $\Lambda$ and $\Sigma$, the other u

**Table 2** The branching ratios of $\Lambda(1115)$ and $\Sigma(1193)$ weak decays from various models, compared with data

| The branching ratio ($\times 10^{-4}$) of $\Lambda(1115)$ | | |
|---|---|---|
| Model | $\bar{\nu}_e + e^- + p$ | $\bar{\nu}_\mu + \mu^- + p$ |
| EXP-A | 8.18 | 1.36 |
| EXP-B | 8.25 | 1.38 |
| R3QM | 9.31 | 1.55 |
| NR3QM-full | 10.7 | 1.82 |
| NR3QM-single | 11.8 | 1.95 |
| Experimental data [14] | $8.32 \pm 0.14$ | $1.57 \pm 0.35$ |
| The branching ratio ($\times 10^{-14}$) of $\Sigma^0(1193)$ | | |
| Model | $\bar{\nu}_e + e^- + p$ | $\bar{\nu}_\mu + \mu^- + p$ |
| EXP | 22.43 | 10.32 |
| R3QM | 20.10 | 9.35 |
| NR3QM-full | 18.82 | 8.82 |
| NR3QM-single | 23.83 | 10.80 |
| Experimental data [14] | - | - |
| The branching ratio ($\times 10^{-4}$) of $\Sigma^-(1193)$ | | |
| Model | $\bar{\nu}_e + e^- + n$ | $\bar{\nu}_\mu + \mu^- + n$ |
| EXP | 9.58 | 4.51 |
| R3QM | 9.30 | 4.06 |
| NR3QM-full | 8.16 | 3.95 |
| NR3QM-single | 10.35 | 4.85 |
| Experimental data [14] | $10.17 \pm 0.34$ | $4.5 \pm 0.4$ |

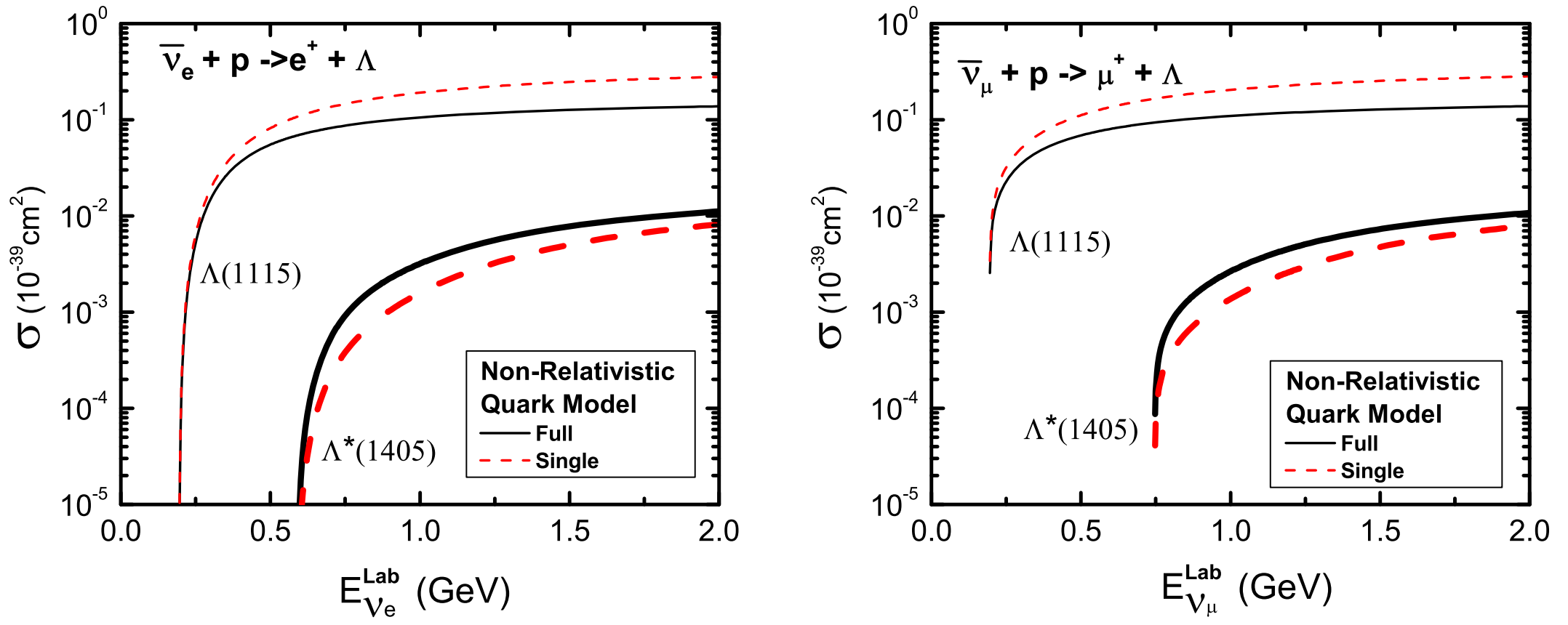

**Fig. 6** NR3QM prediction of total cross sections for $\bar{\nu}_e + p \to \Lambda + e^+$ (*left*) and $\bar{\nu}_\mu + p \to \Lambda + \mu^+$ (*right*) versus neutrino beam energy in the lab system. The *thin and thick lines* are for $\Lambda(1115)$ and $\Lambda^*(1405)$, respectively. The *line types* are defined as the same as in Fig. 4

quark and d quark should keep exchange anti-symmetric and symmetric, respectively. In details, if we assume the third quark is the transition quark in the flavour wave functions of proton and hyperons, for the $\Lambda(1115)$ and $\Sigma(1193)$, only ${}_\rho\langle\Lambda|\chi_s^+\chi_u|p\rangle_\rho = \sqrt{6}/3 \neq 0$ and ${}_\lambda\langle\Sigma|\chi_s^+\chi_u|p\rangle_\lambda = \sqrt{2}/3 \neq 0$, respectively. It leads to the difference between two amplitudes with a factor of $\sqrt{3}$. Furthermore, it also results in the totally different spin transition amplitudes, $p \to \Lambda$: ${}_\rho\langle\frac{1}{2}, s_z^\Lambda|\hat{O}^\nu|\frac{1}{2}, s_z^p\rangle_\rho$ and $p \to \Sigma^0$: ${}_\lambda\langle\frac{1}{2}, s_z^\Sigma|\hat{O}^\nu|\frac{1}{2}, s_z^p\rangle_\lambda$. In "Appendix B", the details of two terms are shown in Eq. (91) and Eqs. (94–97), respectively. If we neglect the contribution from spin flip terms, the ratio between two terms is about 9/5. Combining with the factor $\sqrt{3}$ from the flavour wave function, it naturally explains why the production of $\Lambda(1115)$ is larger than that of $\Sigma(1193)$.

On the other hand, the predicted total cross section for $\Sigma^{0*}(1385)$ is almost same as $\Lambda^*(1405)$. However, as we known, $\Lambda^*(1405)$ may include large five quark components ($uds[u\bar{u}] + uds[d\bar{d}]$) with all quarks in the S-wave. The transition of 3-quark configuration to 5-quark configuration may play an important role for the $\Lambda^*(1405)$ production, as in the case for the photo-production of $N^*(1535)$ [54]. This may make the actual total cross section for the $\Lambda^*(1405)$ production larger than our prediction assuming only 3-quark configuration. In addition, another description of $\Lambda^*(1405)$ is molecular meson-baryon component and a double pole structure [2–6]. We need new wave functions to describe these molecular baryons in quark level, furthermore, these 5-quark components would also increase the total cross section for $\Lambda^*(1405)$. On experimental side, since the events of $\Lambda^*(1405)$ need to be re-constructed from the $\Sigma(1193)\pi$ final state, the non-resonant contribution should be considered in the future.

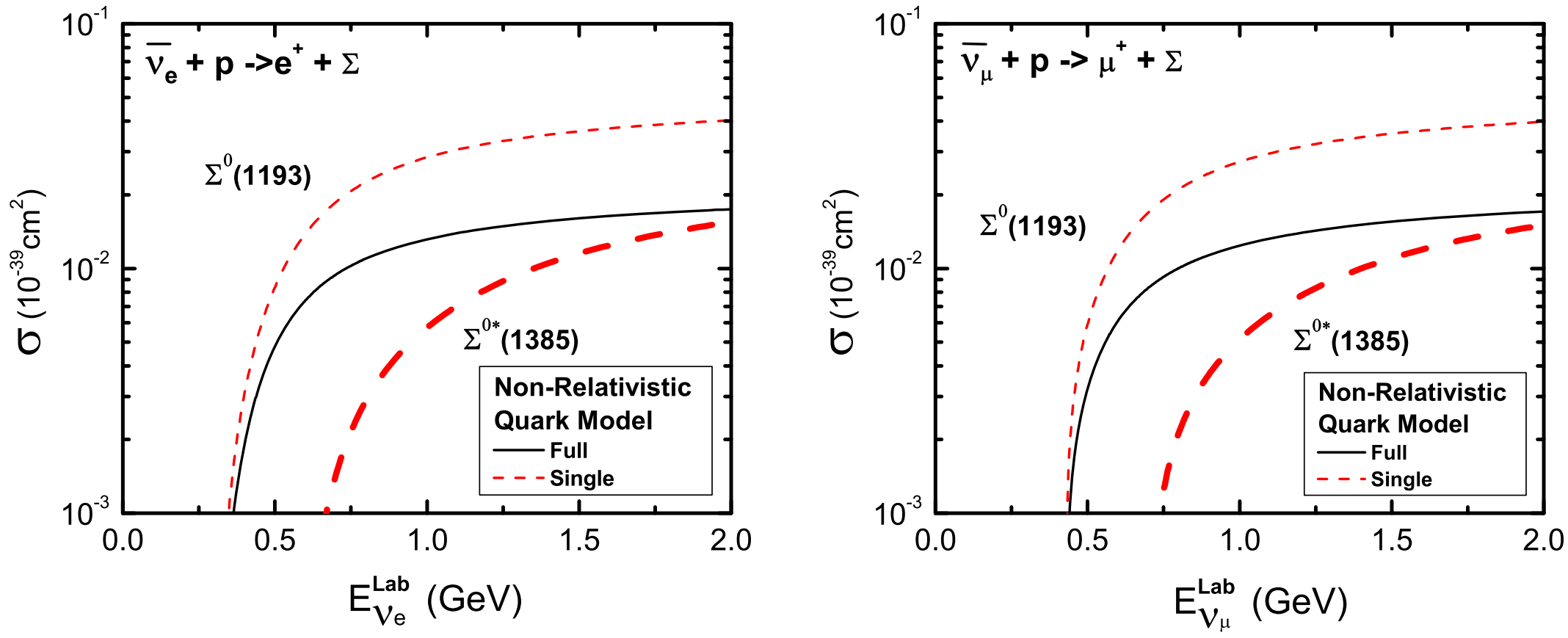

**Fig. 7** NR3QM prediction of total cross sections for $\bar{\nu}_e + p \to \Sigma + e^+$ (*left*) and $\bar{\nu}_\mu + p \to \Sigma + \mu^+$ (*right*) versus neutrino beam energy in the lab system. The *thin and thick lines* are for $\Sigma^0(1193)$ and $\Sigma^{0*}(1385)$, respectively. The *line types* are defined as the same as in Fig. 4

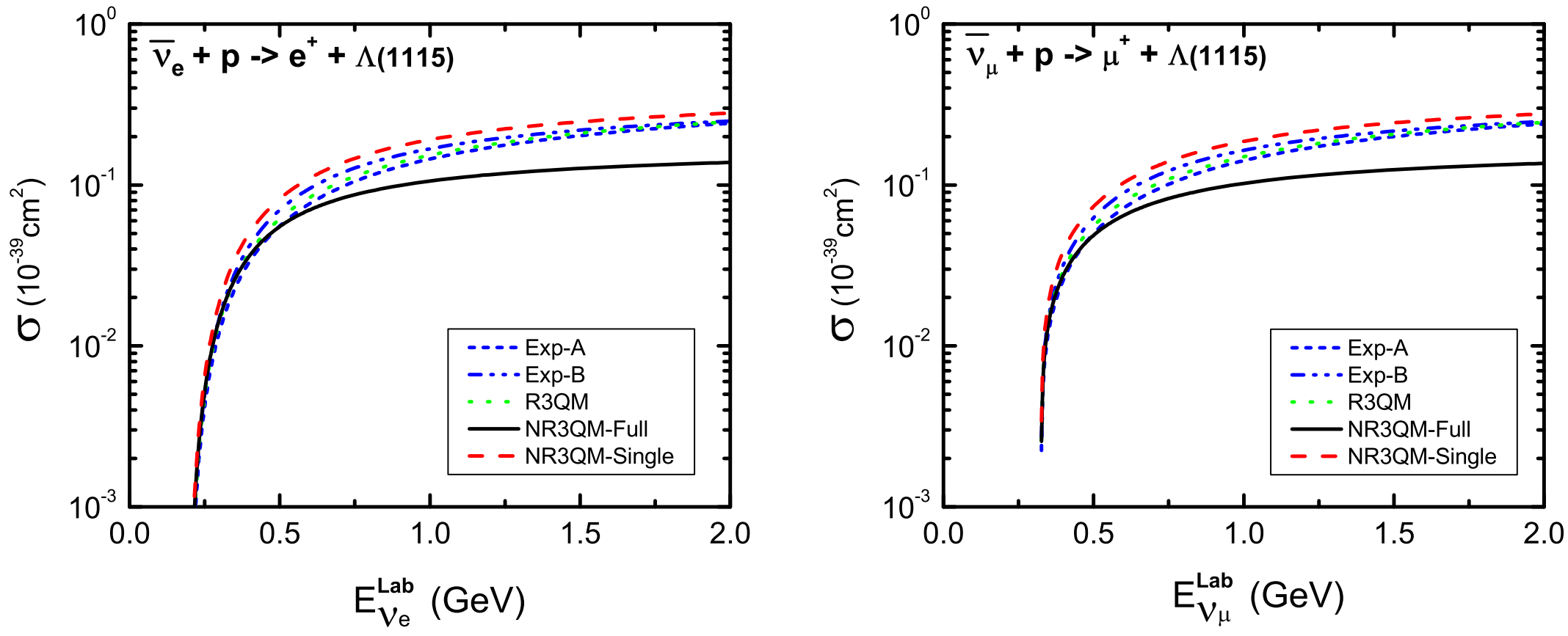

**Fig. 8** (Color online) Comparison of predictions from various models for the total cross section of $\bar{\nu}_e + p \to \Lambda(1115) + e^+$ (*left*) and $\bar{\nu}_\mu + p \to \Lambda(1115) + \mu^+$ (*right*) versus neutrino beam energy in the lab system. The *black solid* and *red dashed lines* are the same as the *thin lines* in Fig. 6. The *blue short-dashed, blue dash-dot-dotted*, and *green dotted lines* are from "Exp-A", "Exp-B", "R3QM" models, respectively

### 3.2 Comparison with V–A Models for $\Lambda/\Sigma$ Weak Decays and Neutrino Production

To further check the reliability of our predictions from the simple non-relativistic 3-quark model, we also use the V–A models to calculate the $\Lambda(1115)/\Sigma(1193)$ weak decays and neutrino production rates. The V–A models with effective form factors either from fitting hyperon weak decay data or from relativistic 3-quark model are used for our calculation.

In Table 2, the branching ratios of $\Lambda(1115)$ and $\Sigma(1193)$ weak decays calculated from various models are listed. The results from the non-relativistic model are larger than those from V–A models, and also 30% larger than experiment data. However, all of the results are consistent in the order of magnitude.

Comparison of predictions from various models for the total cross sections of $\bar{\nu}_{e/\mu}+p \to e^+/\mu^+ + \Lambda(1115)$ and $\bar{\nu}_{e/\mu} + p \to e^+/\mu^+ + \Sigma^0(1193)$ are shown in Figs. 8 and 9, respectively.

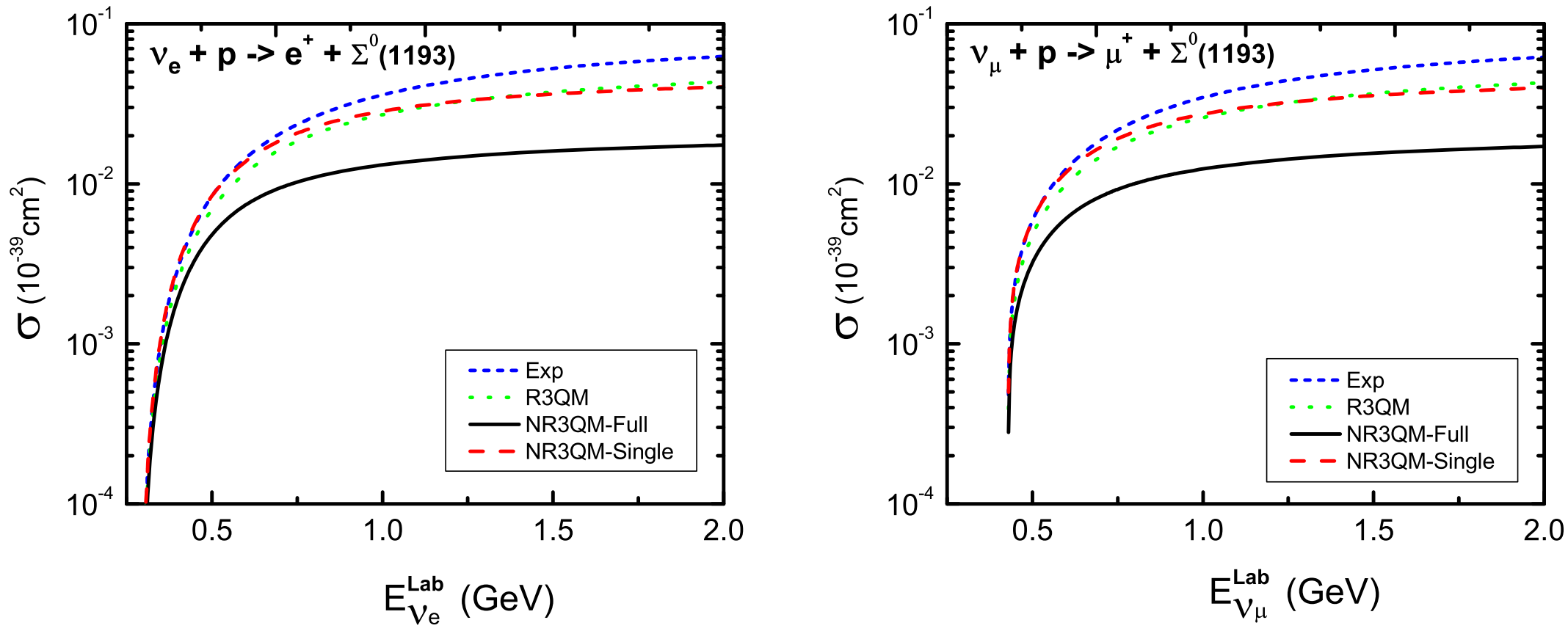

**Fig. 9** Comparison of predictions from various models for the total cross section of $\bar{\nu}_e + p \to \Sigma^0(1193) + e^+$ (*left*) and $\bar{\nu}_\mu + p \to \Sigma^0(1193) + \mu^+$ (*right*) versus neutrino beam energy in the lab system. The *black solid and red dashed lines* are the same as the *thin lines* in Fig. 7. The *blue short-dashed* and *green dotted lines* are from "EXP" and "R3QM" models, respectively

**Table 3** The total cross section ($\times 10^{-40}$ cm$^2$) of $\bar{\nu}_l + p \to \Lambda(1115) + l^+$ and $\bar{\nu}_l + p \to \Sigma^0(1193) + l^+$ at $E_\nu = 2$ GeV in the lab system with three different models, compared with other references' results at the same energy and data with energies shown in the brackets

| | $\bar{\nu}_l + p \to \Lambda(1115) + l^+$ | $\bar{\nu}_l + p \to \Sigma^0(1193) + l^+$ |
|---|---|---|
| EXP | 2.3–2.5 | 0.62 |
| R3QM | 2.4 | 0.43 |
| NR3QM | 1.4–2.8 | 0.18–0.39 |
| Ref. [26] ($l = \mu$) | 1.65 | 0.21 |
| Ref. [27] ($l = \mu$) | 2.0 | 0.65 |
| Ref. [28] ($l = e$) | 5.5–6.5 | 1.1–2.0 |
| Exp [22] ($l = \mu$) | $1.3^{+0.9}_{-0.7}$ (1–3 GeV) | – |
| Exp [23] ($l = \mu$) | $1.17^{+0.41}_{-0.55}$ (0.5–1.5 GeV) | $0.16 \pm 0.10$ (0.5–10 GeV) |
| | $1.84^{+0.59}_{-0.57}$ (1.5–2.5 GeV) | |

In Fig. 8, we see that the different parameters sets (EXP-A and EXP-B), which are from the same experiment [35], give different predictions for the total cross sections of $\bar{\nu}_{e/\mu} + p \to e^+/\mu^+ + \Lambda(1115)$, 0.08 and 0.10 ($\times 10^{-39}$ cm$^2$) at $E_\nu = 0.65$ GeV in the lab system. This means that we can distinguish two sets of parameters in this reaction. Comparing the results from different models, we find that the predictions of NR3QM-Single are a little higher than those from the EXP and R3QM models, while the predictions of NR3QM-Full are smaller than those from other models. In short, the range of total cross section of $\bar{\nu}_{e/\mu} + p \to e^+/\mu^+ + \Lambda(1115)$ from various models is about (0.14–0.28) $\times 10^{-39}$ cm$^2$; the NR3QM-Full gives the lower bound of the range.

In Fig. 9, the estimations of $\Sigma(1193)$ production from the NR3QM-Full are all much smaller than those from the EXP and R3QM by a factor about 2–4, while the NR3QM-Single is comparable with others. At the $E_\nu = 2.0$ GeV in the lab system, the predicted total cross section is 0.62 and 0.43 ($\times 10^{-40}$ cm$^2$) in the EXP and R3QM, respectively, while it is $(0.18 \sim 0.39) \times 10^{-40}$ cm$^2$ in the NR3QM. Furthermore, previous Refs. [26–29] gave the predictions of $\Lambda(1115)$ and $\Sigma^0(1193)$. The comparisons at E = 2 GeV with our results are listed in Table 3 including some experimental data.

The theoretical predictions of $\Sigma(1193)$ from various models suffer a large model uncertainties with a factor of 4 and the experimental data cannot provide any constraints. Thus it is worthy to discuss more about previous Refs. [26–29]. The calculation of Ref. [26] is the generalization of Ref. [55] in the hyperon neutrino production sector. In Ref. [55], they extracted the vector current from a relativistic quark model, while the coupling constant of axial vector current is fitted with experimental data. The form factors of both vector and axial vector currents used the dipole form factor with the same effective mass 0.843 GeV. The transition form factor between resonance and ground state respects the SU(3) symmetry. In principle, this model is similar to our NR3QM without SU(3) breaking effect. Correspondingly, the predictions of Ref. [26] are well consistent with those of NR3QM. The axial vector current is got from the experimental data in Ref. [27], where coupling constant, effective mass and type of form factor are all the same as those in our EXP and EXP-A for $\Lambda(1115)$ case. However, for the vector current part, they use the SU(3) symmetry and the electromagnetic form factor of neutron and proton. As a result, the dependence of the transfer momentum and the coupling constant are both different from our EXP and R3QM. Ref. [28,29] gives predictions by using different types of form factor: the normal dipole form factor and the recoil polarization form factor.

In summary, the predictions from various models are quite consistent with each other for the $\Lambda(1115)$ production, but suffer larger uncertainties for the $\Sigma^0(1193)$ case. Nevertheless, the predictions in all models indicate that the total cross section for the $\Sigma(1193)$ production is much smaller than for the $\Lambda(1115)$ production by a factor about $5\pm2$. In the previous subsection, we have explained why the $\Sigma(1193)$ production is different from $\Lambda(1115)$ production, and by the comparison in this section, it seems that the wave functions of $\Lambda(1115)$ and $\Sigma^0(1193)$ reasonably reflect the nature of these hyperons. The $\Lambda(1115)$ and $\Sigma(1193)$ hyperons are both well established ground states. The dominant components of them are in 3-quark configurations. Naturally, the predictions of NR3QM model are valuable to constrain the theoretical predictions of hyperon production, although our model is too simple to give precise prediction.

For the $\Sigma^{0*}(1385)$ productions, because there is no experimental information available for the vector and axial vector form factors, the vertex of $NW\Sigma^*$ can only be obtained from $NW\Delta$ with SU(3) symmetry [32–34]. However, it is also difficult to get weak current of $NW\Lambda^*(1405)$ by SU(3) symmetry, because there are a lot of uncertainties of $\Lambda^*(1405)$'s structure. So we only use V–A approach to calculate the $\Sigma^{0*}(1385)$ production by the $NW\Sigma^*$ vertex introduced in Ref. [32–34].

The vertex of $p(p)W^-(q)\Sigma^{0*}(1385)(P=p+q)$ is got from Refs. [32–34,50]:

$$T_{NW\Sigma^{0*}(1385)} = \sqrt{\frac{G_F m_W}{\sqrt{2}}}|v_{us}|\bar{u}^{\alpha}_{\Sigma^*(1385)}(V_{\alpha\mu}+A_{\alpha\mu})u_N, \tag{28}$$

$$V_{\alpha\mu} = \left(\frac{C_3^V}{m_N}(g_{\alpha\mu}\not{q}-q_\alpha\gamma_\mu)+\frac{C_4^V}{m_N^2}(g_{\alpha\mu}q\cdot P-q_\alpha P_\mu)\right.$$
$$\left.+\frac{C_5^V}{m_N^2}(g_{\alpha\mu}q\cdot p-q_\alpha p_\mu)+C_6^V g_{\alpha\mu}\right)\gamma^5, \tag{29}$$

$$A_{\alpha\mu} = C_3^A m_N(g_{\alpha\mu}\not{q}-q_\alpha\gamma_\mu)+\frac{C_4^A}{m_N^2}(g_{\alpha\mu}q\cdot P-q_\alpha P_\mu)+C_5^A g_{\alpha\mu}+\frac{C_6^A}{m_N^2}q_\mu q_\alpha, \tag{30}$$

where

$$C_3^V = -\frac{\sqrt{2}}{2}\frac{2.13}{\left(1-\frac{q^2}{m_V^2}\right)^2}\frac{1}{1-\frac{q^2}{4m_V^2}}, \tag{31}$$

$$C_4^V = -\frac{\sqrt{2}}{2}\frac{-1.51}{\left(1-\frac{q^2}{m_V^2}\right)^2}\frac{1}{1-\frac{q^2}{4m_V^2}}, \tag{32}$$

$$C_5^V = -\frac{\sqrt{2}}{2}\frac{0.48}{\left(1-\frac{q^2}{m_V^2}\right)^2}\frac{1}{1-\frac{q^2}{0.776m_V^2}}, \tag{33}$$

$$C_6^V = 0, \tag{34}$$

$$C_3^A = 0, \tag{35}$$

$$C_4^A = -\frac{C_5^A}{4}, \tag{36}$$

$$C_5^A = -\sqrt{\frac{2}{3}}\frac{1}{\left(1-\frac{q^2}{m_A^2}\right)^2}\frac{1}{1-\frac{q^2}{3m_A^2}}, \tag{37}$$

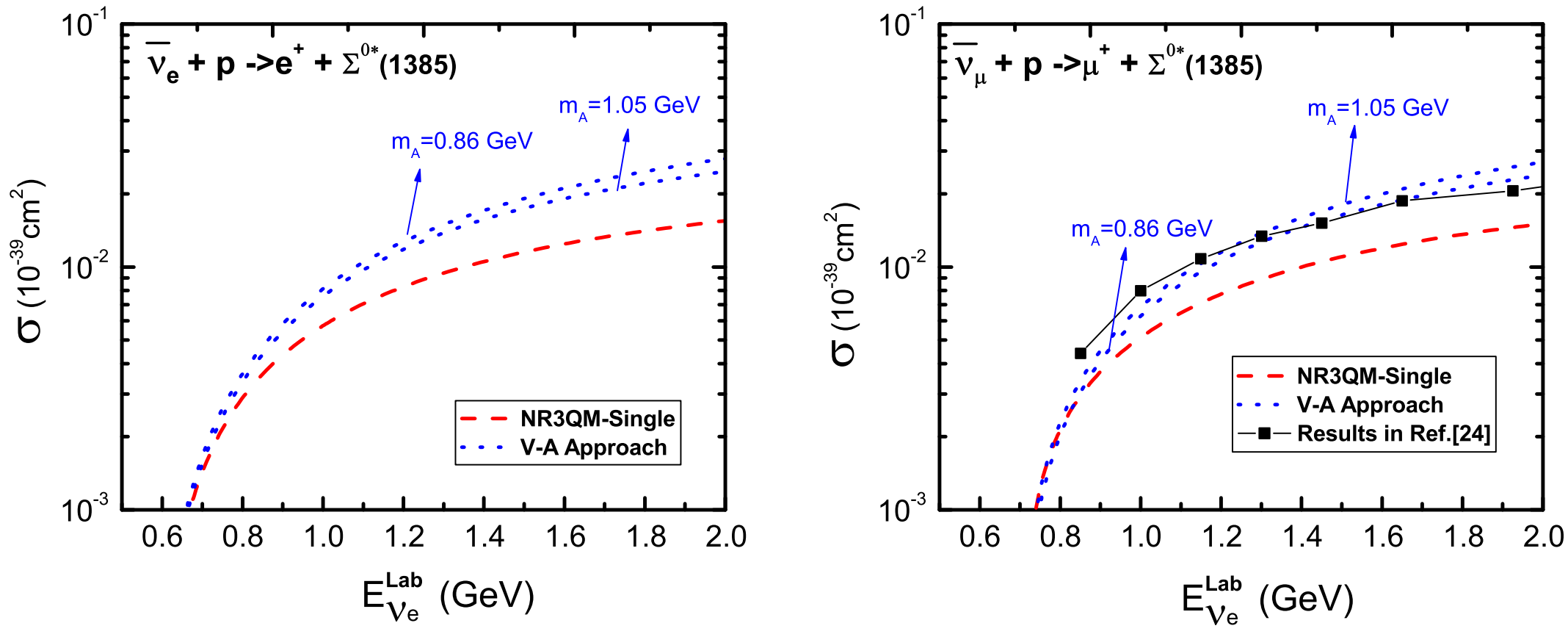

**Fig. 10** (Color online) Various model predictions of total cross sections for $\bar{\nu}_e + p \to \Sigma^{0*}(1385) + e^+$ (*left*) and $\bar{\nu}_\mu + p \to \Sigma^{0*}(1385) + \mu^+$ (*right*) versus neutrino beam energy in the lab system. The *red dashed lines* are defined as the same as those in Fig. 7. The *blue dotted lines* correspond to the V–A approach with different effective mass $m_A$. And the *black square points* are the results in Ref. [26]

$$C_6^A = C_5^A \frac{m_N^2}{m_K^2 - q^2}. \tag{38}$$

The effective masses are $m_V = 0.84\,\text{GeV}$ and $m_A = 1.05\,\text{GeV}$ [50] or $0.86\,\text{GeV}$ [51]. As shown in Fig. 10, the cross section is much more sensitive to the $m_A$ here than that in Fig. 3. It is due to the $J^P$ of hyperon here is $3/2^+$, while it is $1/2^+$ for $\Sigma(1193)$. As a result, two weak current vertices are totally different: here $\gamma_5$ is in the vector current, rather than the axial vector current. Furthermore, the types of the axial vector current form factors are also different: the power of $m_A^2$ is 3 in the $C_5^A$, while it is 2 in Eq. (21). The $m_N$ and $m_K$ are the masses of nucleon and kaon, respectively. The spin wave functions of proton and $\Sigma^*(1385)$ are $u_p$ and $u^\mu_{\Sigma^*(1385)}$, respectively, which satisfy:

$$\sum_{spin} \bar{u}_p u_p = \frac{\not{p} + m_N}{2m_N}, \tag{39}$$

$$\sum_{spin} \bar{u}^\mu_{\Sigma^*} u^\mu_{\Sigma^*} = \frac{\not{P} + m_{\Sigma^*}}{2m_{\Sigma^*}} \left( g^{\mu\nu} - \frac{1}{3}\gamma^\mu\gamma^\nu - \frac{2}{3}\frac{P^\mu P^\nu}{m^2_{\Sigma^*}} + \frac{1}{3}\frac{P^\mu\gamma^\nu - P^\nu\gamma^\mu}{m_{\Sigma^*}} \right) \tag{40}$$

where $m_{\Sigma^*}$ is the mass of $\Sigma^*(1385)$.

With above equations together with Eqs. (1–4), the total cross section of $\bar{\upsilon}_l + p \to l^+ + \Sigma^{0*}(1385)$ can be calculated by the V–A approach. The results are larger than previous predictions with the NR3QM by a factor of 1.5–2 as shown in Fig. 10 which also includes the results of Ref. [26]. Until now, almost no experimental information is available for this reaction. Only Ref. [23] estimates that the production of $\Sigma^*(1385)$ may be comparable to the $\Sigma(1193)$ after some experimental assumptions. Thus, it would be very interesting to measure the exact production rate of $\Sigma^*(1385)$.

From these model calculations, we believe, roughly speaking, the neutrino production rates of $\Lambda(1115)$, $\Lambda^*(1405)$, $\Sigma(1193)$ and $\Sigma^{0*}(1385)$ hyperons are about 1–2 orders of magnitude smaller than the production of $\Delta$. Recently, the MiniBooNE Collaboration reported a collection of about 50,000 candidate $\pi^+$ events with $CH_2$ target at $E_\nu \sim 1\,\text{GeV}$ with an overall signal efficiency of 12.7% [45,46]. This means about a few ten thousands hyperons were also produced in their target. We would suggest the MiniBooNE Collaboration to consider to study hyperon production from their experiment. As the first step, the $\Lambda(1115)$ production should be studied with its $p\pi^-$ decay mode to set up a calibration for possible study of other hyperon resonances. This may open a new window for the study of low energy hyperon resonances in the future accelerator-based neutrino experiments with higher neutrino flux.

## 4 Summary and Outlook

In summary, we use non-relativistic 3-quark model to calculate the total cross sections of $\bar{\nu}_\mu + p \to \mu^+ + n$, $\nu_\mu + n \to \mu^- + p$, and $\nu_\mu + p \to \mu^- + \Delta^{++}$, and the branching ratios of $\Lambda(1115) \to \bar{\nu}_{e/\mu} + e^-/\mu^- + p$ and $\Sigma^-(1193) \to \bar{\nu}_{e/\mu} + e^-/\mu^- + n$. All the data can be reproduced reasonably well with and without SU(3) breaking effect. Then we use this model to predict the cross sections of $\bar{\nu}_{e/\mu} + p \to e^+/\mu^+ + \Sigma^0(1193)$, $\bar{\nu}_{e/\mu} + p \to e^+/\mu^+ + \Sigma^{0*}(1385)$, $\bar{\nu}_{e/\mu} + p \to e^+/\mu^+ + \Lambda(1115)$ and $\bar{\nu}_{e/\mu} + p \to e^+/\mu^+ + \Lambda^*(1405)$ to be about 18–39, ∼10, 140–280 and 8–11 ($\times 10^{-42}$ cm$^2$), respectively, at $E_\nu \sim 2\,\text{GeV}$. Furthermore, we use the vector and axial vector current approach with effective form factors determined either by fitting the $\Lambda(1115)$ and $\Sigma(1193)$ weak decay data or from relativistic quark model to predict the $\Lambda(1115)$ and $\Sigma(1193)$ neutrino production. The results are larger but of the same order of magnitude with the results from the non-relativistic 3-quark model. Based on our predictions in the simple non-relativistic 3-quark model, we estimate that corresponding to about 50,000 candidate $\pi^+$ events collected by the MiniBooNE Collaboration with $CH_2$ target at $E_\nu \sim 1\,\text{GeV}$, there should also be about a few ten thousands hyperons produced in their target. Including more elaborate 5-quark configurations for hyperon resonances may predict higher production rates for them, but not by order of magnitude.

We suggest the MiniBooNE Collaboration and future accelerator-based neutrino experiments with higher neutrino flux to study hyperon production processes. The neutrino induced hyperon production processes $\bar{\nu}_{e/\mu} + p \to e^+/\mu^+ + \pi + \Lambda/\Sigma$ may provide a unique clean place for studying low energy $\pi\Lambda/\Sigma$ interaction and hyperon resonances below $\bar{K}N$ threshold. It would be very helpful for clarifying the puzzling nature of the lowest $J^P = 1/2^-$ hyperon states: $\Lambda^*(1405)$ and possible $\Sigma^*(1390)$.

**Acknowledgments** We thank Wen-Long Zhan, Hu-Shan Xu, T.-S. Harry Lee, T. Sato, Chun-Sheng An and Zi-Du Lin for useful discussions. This work is supported in part by the National Natural Science Foundation of China under Grant 11035006, 11121092, 11261130311 (CRC110 by DFG and NSFC), the Chinese Academy of Sciences under Project No. KJCX2-EW-N01 and the Ministry of Science and Technology of China (2009CB825200). This work is also supported by the U.S. Department of Energy, Office of Nuclear Physics Division, under Contract No. DE-AC02-06CH11357.

## Appendix

### Appendix A: The Baryon's Wave Function

The baryon's wave functions are taken as the same as those in Ref. [9]. For the configurations of the octet baryons, the flavor-spin-orbital wave functions are:

$$\left|B_8^2 S_s, \frac{1}{2}^+\right\rangle = \frac{1}{\sqrt{2}}\left(|B\rangle_\lambda \left|\frac{1}{2}, s_z\right\rangle_\lambda + |B\rangle_\rho \left|\frac{1}{2}, s_z\right\rangle_\rho\right)\Phi_{000}(\mathbf{q}_\lambda, \mathbf{q}_\rho), \tag{41}$$

$$\left|B_8^2 S'_s, \frac{1}{2}^+\right\rangle = \frac{1}{\sqrt{2}}\left(|B\rangle_\lambda \left|\frac{1}{2}, s_z\right\rangle_\lambda + |B\rangle_\rho \left|\frac{1}{2}, s_z\right\rangle_\rho\right)\Phi^s_{200}(\mathbf{q}_\lambda, \mathbf{q}_\rho), \tag{42}$$

$$\begin{aligned}\left|B_8^2 S_M, \frac{1}{2}^+\right\rangle = \frac{1}{2}\Bigg[&\left(|B\rangle_\lambda \left|\frac{1}{2}, s_z\right\rangle_\rho + |B\rangle_\rho \left|\frac{1}{2}, s_z\right\rangle_\lambda\right)\Phi^\rho_{200}(\mathbf{q}_\lambda, \mathbf{q}_\rho) \\ &-\left(|B\rangle_\lambda \left|\frac{1}{2}, s_z\right\rangle_\lambda + |B\rangle_\rho \left|\frac{1}{2}, s_z\right\rangle_\rho\right)\Phi^\lambda_{200}(\mathbf{q}_\lambda, \mathbf{q}_\rho)\Bigg].\end{aligned} \tag{43}$$

For the baryon-decuplet $\Delta^{++}$ and $\Sigma^{0*}(1385)$, their flavor-spin-orbital wave functions are:

$$\left|\Delta^{++} S, \frac{3}{2}^+\right\rangle = \left|\Delta^{++}\right\rangle \left|\frac{3}{2}, s_z\right\rangle \Phi_{000}(\mathbf{q}_\lambda, \mathbf{q}_\rho), \tag{44}$$

$$\left|\Sigma^{0*}(1385) S, \frac{3}{2}^+\right\rangle = \left|\Sigma^{0*}(1385)\right\rangle \left|\frac{3}{2}, s_z\right\rangle \Phi_{000}(\mathbf{q}_\lambda, \mathbf{q}_\rho). \tag{45}$$

For the configurations of $\Lambda^*(1405)$, the flavor-spin-orbital wave functions are:

$$\left|\Lambda_1^2 P_A, \frac{1}{2}^-\right\rangle = \frac{1}{\sqrt{6}} \left|\Lambda\right\rangle_a \; X_a \Phi_\Lambda(\mathbf{q}_\lambda, \mathbf{q}_\rho), \tag{46}$$

$$\left|\Lambda_8^2 P_M, \frac{1}{2}^-\right\rangle = -\frac{1}{2\sqrt{3}} \left(\left|\Lambda\right\rangle_\lambda \; X_\lambda + \left|\Lambda\right\rangle_\rho \; X_\rho\right) \Phi_\Lambda(\mathbf{q}_\lambda, \mathbf{q}_\rho), \tag{47}$$

$$\left|\Lambda_8^4 P_M, \frac{1}{2}^-\right\rangle = \frac{1}{2\sqrt{3}} \left(\left|\Lambda\right\rangle_\lambda \; X'_\lambda + \left|\Lambda\right\rangle_\rho \; X'_\rho\right) \Phi_\Lambda(\mathbf{q}_\lambda, \mathbf{q}_\rho). \tag{48}$$

In these flavor-spin-orbital wave functions, $|B\rangle_{\lambda/\rho}$, $|\Lambda\rangle_{a/\lambda/\rho}$, and $|\Delta^{++}\rangle$ are the flavor wave functions; $X_a$, $X_{\lambda/\rho}$, and $X'_{\lambda/\rho}$ are the spin-orbital coupled wave functions; $|\frac{1}{2}, s_z\rangle_{\lambda/\rho}$ and $|\frac{3}{2}, s_z\rangle$ are the spin wave functions. All of these are used to construct $|X^Q, s_z^Q\rangle$ in Eq. (16). $\Phi_{000}(\mathbf{q}_\lambda, \mathbf{q}_\rho)$, $\Phi^s_{200}(\mathbf{q}_\lambda, \mathbf{q}_\rho)$, $\Phi^\rho_{200}(\mathbf{q}_\lambda, \mathbf{q}_\rho)$, $\Phi^\rho_{200}(\mathbf{q}_\lambda, \mathbf{q}_\rho)$, and $\Phi_\Lambda(\mathbf{q}_\lambda, \mathbf{q}_\rho)$ denote the orbital wave functions; and the Jacobin momenta are related to those of the quarks by

$$\mathbf{q}_\rho = \frac{\mathbf{q}_1 - \mathbf{q}_2}{\sqrt{2}}, \tag{49}$$

$$\mathbf{q}_\lambda = \frac{\mathbf{q}_1 + \mathbf{q}_2 - 2\mathbf{q}_3}{\sqrt{6}}. \tag{50}$$

The involved flavor wave functions, spin wave functions and orbital wave functions are given explicitly as the following.
The flavor wave functions:

$$|\Lambda\rangle_a = \frac{1}{\sqrt{6}} \left(|uds\rangle + |dsu\rangle + |sud\rangle - |usd\rangle - |sdu\rangle - |dus\rangle\right), \tag{51}$$

$$|\Lambda\rangle_\rho = \frac{1}{2\sqrt{3}} \left(|usd\rangle - |dsu\rangle - |sud\rangle + |sdu\rangle + 2\,|uds\rangle - 2\,|dus\rangle\right), \tag{52}$$

$$|\Lambda\rangle_\lambda = \frac{1}{2} \left(|usd\rangle + |sud\rangle - |sdu\rangle - |dsu\rangle\right), \tag{53}$$

$$\left|\Sigma^0\right\rangle_\rho = \frac{1}{2} \left(|usd\rangle + |dsu\rangle - |sud\rangle - |sdu\rangle\right), \tag{54}$$

$$\left|\Sigma^0\right\rangle_\lambda = -\frac{1}{2\sqrt{3}} \left(|usd\rangle + |dsu\rangle + |sud\rangle + |sdu\rangle - 2\,|uds\rangle - 2\,|dus\rangle\right), \tag{55}$$

$$\left|\Sigma^-\right\rangle_\rho = \frac{1}{\sqrt{2}} \left(|dsd\rangle - |sdd\rangle\right), \tag{56}$$

$$\left|\Sigma^-\right\rangle_\lambda = \frac{1}{6} \left(2\,|dds\rangle - |sdd\rangle - |dsd\rangle\right), \tag{57}$$

$$|p\rangle_\lambda = \frac{1}{\sqrt{6}} \left(2\,|uud\rangle - |duu\rangle - |udu\rangle\right), \tag{58}$$

$$|p\rangle_\rho = \frac{1}{\sqrt{2}} \left(|udu\rangle - |duu\rangle\right), \tag{59}$$

$$|n\rangle_\lambda = -\frac{1}{\sqrt{6}} \left(2\,|ddu\rangle - |udd\rangle - |dud\rangle\right), \tag{60}$$

$$|n\rangle_\rho = -\frac{1}{\sqrt{2}} \left(|dud\rangle - |udd\rangle\right), \tag{61}$$

$$\left|\Delta^{++}\right\rangle = |uuu\rangle, \tag{62}$$

$$\left|\Sigma^{0*}(1385)\right\rangle = \frac{1}{\sqrt{6}} \left(|uds\rangle + |dus\rangle + |dsu\rangle + |sdu\rangle + |sud\rangle + |usd\rangle\right). \tag{63}$$

The spin-orbital coupled wave functions [7] ($X^{\pm}$ for $s_z$ = 1/2 and −1/2):

$$X_a^+ = -\left|\frac{1}{2},\frac{1}{2}\right\rangle_\lambda (\rho,0) + \sqrt{2}\left|\frac{1}{2},-\frac{1}{2}\right\rangle_\lambda (\rho,+1) + \left|\frac{1}{2},\frac{1}{2}\right\rangle_\rho (\lambda,0) - \sqrt{2}\left|\frac{1}{2},-\frac{1}{2}\right\rangle_\rho (\lambda,+1), \quad (64)$$

$$X_a^- = \left|\frac{1}{2},-\frac{1}{2}\right\rangle_\lambda (\rho,0) - \sqrt{2}\left|\frac{1}{2},\frac{1}{2}\right\rangle_\lambda (\rho,-1) - \left|\frac{1}{2},-\frac{1}{2}\right\rangle_\rho (\lambda,0) + \sqrt{2}\left|\frac{1}{2},\frac{1}{2}\right\rangle_\rho (\lambda,+1), \quad (65)$$

$$X_\lambda^+ = -\left|\frac{1}{2},\frac{1}{2}\right\rangle_\lambda (\lambda,0) + \sqrt{2}\left|\frac{1}{2},-\frac{1}{2}\right\rangle_\lambda (\lambda,+1) + \left|\frac{1}{2},\frac{1}{2}\right\rangle_\rho (\rho,0) - \sqrt{2}\left|\frac{1}{2},-\frac{1}{2}\right\rangle_\rho (\rho,+1), \quad (66)$$

$$X_\lambda^- = \left|\frac{1}{2},-\frac{1}{2}\right\rangle_\lambda (\lambda,0) - \sqrt{2}\left|\frac{1}{2},\frac{1}{2}\right\rangle_\lambda (\lambda,-1) - \left|\frac{1}{2},-\frac{1}{2}\right\rangle_\rho (\rho,0) + \sqrt{2}\left|\frac{1}{2},\frac{1}{2}\right\rangle_\rho (\rho,-1), \quad (67)$$

$$X_\rho^+ = \left|\frac{1}{2},\frac{1}{2}\right\rangle_\lambda (\rho,0) - \sqrt{2}\left|\frac{1}{2},-\frac{1}{2}\right\rangle_\lambda (\rho,+1) + \left|\frac{1}{2},\frac{1}{2}\right\rangle_\rho (\lambda,0) - \sqrt{2}\left|\frac{1}{2},-\frac{1}{2}\right\rangle_\rho (\lambda,+1), \quad (68)$$

$$X_\rho^- = -\left|\frac{1}{2},-\frac{1}{2}\right\rangle_\lambda (\rho,0) + \sqrt{2}\left|\frac{1}{2},\frac{1}{2}\right\rangle_\lambda (\rho,-1) - \left|\frac{1}{2},-\frac{1}{2}\right\rangle_\rho (\lambda,0) + \sqrt{2}\left|\frac{1}{2},\frac{1}{2}\right\rangle_\rho (\lambda,-1), \quad (69)$$

$$X_\lambda^{\prime +} = \sqrt{3}\left|\frac{3}{2},\frac{3}{2}\right\rangle(\lambda,-1) - \sqrt{2}\left|\frac{3}{2},\frac{1}{2}\right\rangle(\lambda,0) + \left|\frac{3}{2},-\frac{1}{2}\right\rangle(\lambda,1), \quad (70)$$

$$X_\lambda^{\prime -} = \sqrt{3}\left|\frac{3}{2},-\frac{3}{2}\right\rangle(\lambda,+1) - \sqrt{2}\left|\frac{3}{2},-\frac{1}{2}\right\rangle(\lambda,0) + \left|\frac{3}{2},\frac{1}{2}\right\rangle(\lambda,-1), \quad (71)$$

$$X_\rho^{\prime +} = \sqrt{3}\left|\frac{3}{2},\frac{3}{2}\right\rangle(\rho,-1) - \sqrt{2}\left|\frac{3}{2},\frac{1}{2}\right\rangle(\rho,0) + \left|\frac{3}{2},-\frac{1}{2}\right\rangle(\rho,1), \quad (72)$$

$$X_\rho^{\prime -} = \sqrt{3}\left|\frac{3}{2},-\frac{3}{2}\right\rangle(\rho,+1) - \sqrt{2}\left|\frac{3}{2},-\frac{1}{2}\right\rangle(\rho,0) + \left|\frac{3}{2},\frac{1}{2}\right\rangle(\rho,-1), \quad (73)$$

with $(\lambda,0) = q_{\lambda,z}$, $(\lambda,\pm 1) = \mp\frac{1}{\sqrt{2}}(q_{\lambda,x} \pm i q_{\lambda,y}$, $(\rho,0) = q_{\rho,z}$, and $(\rho,\pm 1) = \mp\frac{1}{\sqrt{2}}(q_{\rho,x} \pm i q_{\rho,y})$.
The spin wave functions:

$$\left|\frac{1}{2},+\frac{1}{2}\right\rangle_\rho = \frac{1}{\sqrt{2}}(|\uparrow\downarrow\uparrow\rangle - |\downarrow\uparrow\uparrow\rangle), \quad (74)$$

$$\left|\frac{1}{2},-\frac{1}{2}\right\rangle_\rho = \frac{1}{\sqrt{2}}(|\uparrow\downarrow\downarrow\rangle - |\downarrow\uparrow\downarrow\rangle), \quad (75)$$

$$\left|\frac{1}{2},+\frac{1}{2}\right\rangle_\lambda = \frac{1}{\sqrt{6}}(2\,|\uparrow\uparrow\downarrow\rangle - |\uparrow\downarrow\uparrow\rangle - |\downarrow\uparrow\uparrow\rangle), \quad (76)$$

$$\left|\frac{1}{2},-\frac{1}{2}\right\rangle_\lambda = -\frac{1}{\sqrt{6}}(2\,|\downarrow\downarrow\uparrow\rangle - |\downarrow\uparrow\downarrow\rangle - |\uparrow\downarrow\downarrow\rangle), \quad (77)$$

$$\left|\frac{3}{2},+\frac{3}{2}\right\rangle = |\uparrow\uparrow\uparrow\rangle\,, \tag{78}$$

$$\left|\frac{3}{2},+\frac{1}{2}\right\rangle = \frac{1}{\sqrt{3}}(|\uparrow\downarrow\uparrow\rangle + |\downarrow\uparrow\uparrow\rangle + |\uparrow\uparrow\downarrow\rangle), \tag{79}$$

$$\left|\frac{3}{2},-\frac{1}{2}\right\rangle = \frac{1}{\sqrt{3}}(|\uparrow\downarrow\uparrow\rangle + |\downarrow\uparrow\uparrow\rangle + |\uparrow\uparrow\downarrow\rangle), \tag{80}$$

$$\left|\frac{3}{2},-\frac{3}{2}\right\rangle = |\downarrow\downarrow\downarrow\rangle\,. \tag{81}$$

The orbital wave functions [9]:

$$\Phi_{000}(\mathbf{q}_\lambda,\mathbf{q}_\rho) = \frac{1}{(\sqrt{\pi}\omega_3)^3}\exp\left\{-\frac{\mathbf{q}_\lambda^2+\mathbf{q}_\rho^2}{2\omega_3^2}\right\}, \tag{82}$$

$$\Phi^s_{200}(\mathbf{q}_\lambda,\mathbf{q}_\rho) = \frac{1}{\sqrt{3}(\sqrt{\pi}\omega_3)^3}\left(3-\frac{\mathbf{q}_\lambda^2+\mathbf{q}_\rho^2}{\omega_3^2}\right)\exp\left\{-\frac{\mathbf{q}_\lambda^2+\mathbf{q}_\rho^2}{2\omega_3^2}\right\}, \tag{83}$$

$$\Phi^\rho_{200}(\mathbf{q}_\lambda,\mathbf{q}_\rho) = \frac{2}{\sqrt{3}\omega_3^2(\sqrt{\pi}\omega_3)^3}\left(\mathbf{q}_\lambda\cdot\mathbf{q}_\rho\right)\exp\left\{-\frac{\mathbf{q}_\lambda^2+\mathbf{q}_\rho^2}{2\omega_3^2}\right\}, \tag{84}$$

$$\Phi^\lambda_{200}(\mathbf{q}_\lambda,\mathbf{q}_\rho) = \frac{1}{\sqrt{3}\omega_3^2(\sqrt{\pi}\omega_3)^3}\left(\mathbf{q}_\rho^2-\mathbf{q}_\lambda^2\right)\exp\left\{-\frac{\mathbf{q}_\lambda^2+\mathbf{q}_\rho^2}{2\omega_3^2}\right\}, \tag{85}$$

$$\Phi_{\Lambda^*}(\mathbf{q}_\lambda,\mathbf{q}_\rho) = \frac{\sqrt{2}}{\sqrt{\pi}^3\omega_3^4}\exp\left\{-\frac{\mathbf{q}_\lambda^2+\mathbf{q}_\rho^2}{2\omega_3^2}\right\}. \tag{86}$$

## Appendix B: Nucleon-Baryon-W Boson Vertices $T_2^\nu$

In this appendix, we explain how to calculate $T_2^\nu$ using baryon's wave functions by taking $\Lambda(1115)$-proton-W boson coupling as an example.
For the flavor part:

$$_\lambda\left\langle\Lambda|\chi_s^+\chi_u|p\right\rangle_\lambda = 0, \tag{87}$$

$$_\lambda\left\langle\Lambda|\chi_s^+\chi_u|p\right\rangle_\rho = 0, \tag{88}$$

$$_\rho\left\langle\Lambda|\chi_s^+\chi_u|p\right\rangle_\lambda = 0, \tag{89}$$

$$_\rho\left\langle\Lambda|\chi_s^+\chi_u|p\right\rangle_\rho = \frac{\sqrt{6}}{3}. \tag{90}$$

For the spin part:

$$_\rho\left\langle\frac{1}{2},s_z^\Lambda|\hat{O}^\mu|\frac{1}{2},s_z^p\right\rangle_\rho = O^\mu(s_z^\Lambda,s_z^p), \tag{91}$$

$$_\rho\left\langle\frac{1}{2},s_z^\Lambda|\hat{O}^\mu|\frac{1}{2},s_z^p\right\rangle_\lambda = 0, \tag{92}$$

$$_\lambda\left\langle\frac{1}{2},s_z^\Lambda|\hat{O}^\mu|\frac{1}{2},s_z^p\right\rangle_\rho = 0, \tag{93}$$

$$_\lambda\left\langle\frac{1}{2},s_z^\Lambda=\frac{1}{2}|\hat{O}^\mu|\frac{1}{2},s_z^p=\frac{1}{2}\right\rangle_\lambda = \frac{1}{3}\left(O^\mu\left(\frac{1}{2},\frac{1}{2}\right)+2O^\mu\left(-\frac{1}{2},-\frac{1}{2}\right)\right), \tag{94}$$

$$_\lambda\left\langle\frac{1}{2},s_z^\Lambda=-\frac{1}{2}|\hat{O}^\mu|\frac{1}{2},s_z^p=\frac{1}{2}\right\rangle_\lambda = \frac{1}{3}O^\mu\left(-\frac{1}{2},\frac{1}{2}\right), \tag{95}$$

$$ {}_{\lambda}\left\langle\frac{1}{2}, s_z^{\Lambda}=\frac{1}{2}|\hat{O}^{\mu}|\frac{1}{2}, s_z^{p}=-\frac{1}{2}\right\rangle_{\lambda}=\frac{1}{3}O^{\mu}\left(\frac{1}{2},-\frac{1}{2}\right), \quad (96) $$

$$ {}_{\lambda}\left\langle\frac{1}{2}, s_z^{\Lambda}=-\frac{1}{2}|\hat{O}^{\mu}|\frac{1}{2}, s_z^{p}=-\frac{1}{2}\right\rangle_{\lambda}=\frac{1}{3}\left(2O^{\mu}\left(\frac{1}{2},\frac{1}{2}\right)+O^{\mu}\left(-\frac{1}{2},-\frac{1}{2}\right)\right), \quad (97) $$

where the vertex $O^{\mu}(s_z^s, s_z^u)$ is:

$$ O^{\mu}\left(s_z^s, s_z^u\right)=\bar{u}_s\left(\mathbf{q}_s,\ s_z^s\right)\gamma^{\nu}\left(1-\gamma^5\right)u_u\left(\mathbf{q}_u,\ s_z^u\right) \quad (98) $$

By using Eq. (12–16), and above equations, we can get $T^{\mu}_{2\ \Lambda-p-W}$ as:

$$ \begin{aligned} T^{\mu}_{2\ \Lambda-p-W}(s_z^{\Lambda}, s_z^{p}) = \int -\frac{9}{2}d\mathbf{q}_u d\mathbf{q}_{\rho}\sqrt{\left(\frac{G_F m_W^2}{\sqrt{2}}\right)}|v_{us}| \\ \times\left\{\left[\frac{0.90}{\sqrt{2}}\Phi_{000}(\mathbf{q}_{\lambda}^{p}, \mathbf{q}_{\rho})+\frac{0.34}{\sqrt{2}}\Phi^{s}_{200}(\mathbf{q}_{\lambda}^{p}, \mathbf{q}_{\rho})+\frac{0.27}{2}\Phi^{\lambda}_{200}(\mathbf{q}_{\lambda}^{p}, \mathbf{q}_{\rho})\right]\right. \\ \times\left[\frac{0.93}{\sqrt{2}}\Phi_{000}(\mathbf{q}_{\lambda}^{\Lambda}, \mathbf{q}_{\rho})+\frac{0.30}{\sqrt{2}}\Phi^{s}_{200}(\mathbf{q}_{\lambda}^{\Lambda}, \mathbf{q}_{\rho})+\frac{0.20}{2}\Phi^{\lambda}_{200}(\mathbf{q}_{\lambda}^{\Lambda}, \mathbf{q}_{\rho})\right] \\ \times {}_{\rho}\langle\frac{1}{2}, s_z^{\Lambda}|\hat{O}^{\nu}|\frac{1}{2}, s_z^{p}\rangle_{\rho} \\ \left.+\frac{0.27}{2}\frac{0.20}{2}\Phi^{\lambda}_{200}(\mathbf{q}_{\lambda}^{p}, \mathbf{q}_{\rho})\Phi^{\lambda}_{200}(\mathbf{q}_{\lambda}^{\Lambda}, \mathbf{q}_{\rho})_{\lambda}\langle\frac{1}{2}, s_z^{\Lambda}|\hat{O}^{\nu}|\frac{1}{2}, s_z^{p}\rangle_{\lambda}\right\} \end{aligned} \quad (99) $$

where $\mathbf{q}_{\lambda}^{p}=-3\mathbf{q}_u/\sqrt{6}$, $\mathbf{q}_{\lambda}^{\Lambda}=-(3\mathbf{q}_u+2\mathbf{q}_W)/\sqrt{6}$.